\documentclass[twocolumn,showpacs,preprintnumbers,amsmath,amssymb]{revtex4}

\usepackage{graphicx}
\usepackage{dcolumn}
\usepackage{bm}
\usepackage{lmodern}

\begin{document}

\title{Modification of the Brink-Axel Hypothesis for High Temperature Nuclear Weak Interactions}

\author{G. Wendell Misch$^{1}$}
\author{George M. Fuller$^{1}$}
\author{B. Alex Brown$^{2}$}
\affiliation{$^{1}$Department of Physics, University of California, San Diego, La Jolla, CA 92093, USA}
\affiliation{$^{2}$National Superconducting Cyclotron Laboratory, and Department of Physics and Astronomy, Michigan State University, East Lansing, Michigan 48824, USA}

\date{\today}

\begin{abstract}
We present shell model calculations of electron capture strength distributions in A=28 nuclei and computations of the corresponding capture rates in supernova core conditions.  We find that in these nuclei the Brink-Axel hypothesis for the distribution of Gamow-Teller strength fails at low and moderate initial excitation energy, but may be a valid tool at high excitation.  The redistribution of GT strength at high initial excitation may affect capture rates during collapse.  If these trends which we have found in lighter nuclei also apply for the heavier nuclei which provide the principal channels for neutronization during stellar collapse, then there could be two implications for supernova core electron capture physics.  First, a modified Brink-Axel hypothesis could be a valid approximation for use in collapse codes.  Second, the electron capture strength may be moved down significantly in transition energy, which would likely have the effect of increasing the overall electron capture rate during stellar collapse.
\end{abstract}

\maketitle

\section{Introduction}
\label{sec:introduction}
The Brink-Axel hypothesis posits that the electromagnetic giant dipole resonance in nuclei resides at the same relative energy from excited states as it does from the ground state \cite{brink:thesis,axel:1962}.  That is, if a given nucleus in its ground state has the resonance at 10 MeV excitation, then that same nucleus in an excited state would have that resonance at 10 MeV above the excited level, and indeed experiment bears this out \cite{szeflinski-etal:1983}.  The idea that the response properties of excited states might be similar to those of the ground state is alluring, especially for astrophysical weak interaction calculations, in part because it can be difficult to measure or calculate excited state properties and relatively easier to study ground state properties.  Moreover, in stellar collapse environments, nuclei can reside in highly excited states, and an approximation like Brink-Axel for Gamow-Teller (we will frequently abbreviate this to GT) transitions is widely applied \cite{ffn:1980,afwh:1994,krs:1994,ckrs:1999,kajino-etal:1988,lm:1998,clmn:1999,lm:1999,ml:1999,lm:2000,lm:2001,cole-etal:2012,flm:2013,lkd:2001,martinez-pinedo-etal:2014,juodagalvis-etal:2010}.  (In fact, for isovector Fermi transitions, the Brink-Axel hypothesis holds exactly insofar as isospin is a good quantum number of the nucleus.)  In this paper we examine electron capture strength on nuclei with high initial excitation energy and its effect on the electron capture rate, with a particular emphasis on adapting the Brink-Axel hypothesis for use in this channel.

Neutronization of the collapsing core through electron capture is pivotally important in the supernova problem, as electrons provide pressure support within the core.  During infall, the mass of the homologous inner core (that portion which collapses subsonically) is set by the electron-to-baryon ratio Y$_e$.  This mass, which acts as a sort of piston at core bounce, sets the initial post-bounce shock energy.  Moreover, Y$_e$ figures into the nuclear composition of the outer core, which dissipates much of the shock energy through photodissociation of its nuclei and affects neutrino transport through coherent interaction with nuclei \cite{arnett:1977,bw:1982,bw:1985,bmd:2003,bm:2007,sjfk:2008,bbol:2011,ajs:2007,liebendorfer-etal:2008,liebendorfer-etal:2009,hjm:2010,hix-etal:2010,bdm:2012}.

During supernova core collapse, the density is very high, starting at around $10^{10}$ g/cm$^3$ at the onset of collapse and proceeding to $>$$10^{14}$ g/cm$^3$ at bounce.  The temperature is very high at $\sim$1-2 MeV, but the entropy per baryon is extremely low at $\approx$1 unit of Boltzmann's constant per baryon \cite{bbal:1979}.  Although electrons are most readily captured onto free protons, the low entropy favors large nuclei which are then in turn the principal sites for electron capture \cite{bbal:1979,arnett:1977,fuller:1982}.  The core is initially cooled during collapse by neutrino emission \cite{btz:1974,dkst:1976,fm:1991,mbf:2013}, so the entropy remains low.  Furthermore, the high temperature puts the nuclei into extremely highly excited states.  Using the Bethe approximation for nuclear level density \cite{bethe:1936}, the average excitation energy is
\begin{equation}
E\approx a(k_BT)^2
\end{equation}
where a $\approx\frac{A}{8}$ MeV$^{-1}$ is the level density parameter.  With a typical nuclear mass of $\sim$120, we find the average excitation energy to be between 15 and 60 MeV.  Finally, as the collapse progresses, the core electron fraction tends toward Y$_e \approx 0.32$, which implies neutron rich nuclei.  In order to understand neutronization during core collapse, we must therefore consider the capture of electrons onto large, highly excited, and eventually neutron rich nuclei.

Large, highly excited, neutron rich nuclei are, unfortunately, problematic to understand both experimentally and theoretically.  Experimental data on these nuclei is sparse \cite{toi}, and while large nuclei certainly exist in abundance, there are as of yet no experimental means by which to put them into high energy states without utterly destroying them.  The (n,p), (p,n), ($^3$He,t), (d,$^2$He), and similar charge exchange channels give information on the Gamow-Teller structure \cite{yako-etal:2005,yako-etal:2009,fpte:2013}, but these experiments can only probe nuclei in the ground state, whereas low entropy, high temperature environments favor much higher excitations.  The Extreme Light Infrastructure may eventually be able to provide some insight into the structure and behavior of highly excited nuclei through the use of multiple MeV laser light, but it is not yet in operation \cite{eli}.  Of course, even when high energy states become readily attainable, we still face the problem that nuclei of the appropriate neutron richness are highly unstable in the laboratory; it is the high density and low entropy of the supernova core that allows them to exist in that environment.

From the theoretical direction, we should look for trends in the Gamow-Teller electron capture strength distribution, as the Brink-Axel hypothesis has had experimental success in the electromagnetic channel.  Fuller, Fowler, and Newman \cite{ffn:1980,ffn:1982a,ffn:1982b,ffn:1985} (hereafter FFNI, FFNII, FFNIII, and FFNIV, respectively, for those specific papers, and FFN for the body of work as a whole) was the first to adopt the Brink-Axel hypothesis for use in the Gamow-Teller charged current channel (we will call this and similar techniques the GT Brink-Axel hypothesis to distinguish it from the experimentally verified giant dipole electromagnetic phenomenon).  This approach and modifications thereof have since been widely used to compute weak rates.  Variations include essentially copying the FFN approach \cite{afwh:1994}, using a broad GT resonance that is the same for all excited states \cite{krs:1994,ckrs:1999}, computing in detail only the lowest few states in the parent and/or daughter nuclei and employing the GT Brink-Axel hypothesis to treat the bulk of the strength at high excitations or neglecting highly excited states entirely \cite{kajino-etal:1988,lm:1998,clmn:1999,lm:1999,ml:1999,lm:2000,lm:2001,cole-etal:2012,flm:2013}, and using thermal averaging techniques \cite{lkd:2001,martinez-pinedo-etal:2014}.  Recently, electron capture rates have been tabulated using combinations of these approaches over a wide range of nuclear masses and stellar conditions \cite{juodagalvis-etal:2010}.

However, there is mounting evidence that we would be unwise to take the Brink-Axel hypothesis at face value.  Angell et al \cite{angell-etal:2012} have shown experimentally that Brink-Axel does not hold for the pygmy dipole resonance, and Nabi \& Sajjad \cite{ns:2007,nabi:2011,nabi:2012} have observed in their theoretical calculations the failure of Brink-Axel for the Gamow-Teller interaction even at modest excitation energies.  Thus, whenever it is computationally feasible, we should avoid use of the GT Brink-Axel hypothesis.  Oda et al \cite{oda-etal:1994} performed full sd shell model computations of the first 100 excited states in each sd shell nucleus, while others have taken to the random phase approximation to examine heavier nuclei \cite{dzhioev-etal:2010,sarriguren:2013}.  But the Oda et al approach of neglecting states higher than the 100$^{th}$ excitation may miss some important features of higher-lying states, and while RPA does well at determining the overall strength distribution, it is unable to accurately reproduce the detailed distributions to which electron capture rates are sensitive.   We are therefore well served by scrutinizing detailed strength distributions up to very high initial excitation to learn in what ways the distribution evolves.  We will show that at least in the sd shell, a modified form of the GT Brink-Axel hypothesis derived from large scale shell model calculations can both be computationally tractable and capture features of the strength distribution at low and high excitation with consequences for core collapse.

Computationally, large nuclei are difficult to study simply because of the large number of nucleons involved; the sheer combinatorics of so many nucleons rapidly drives up the computational requirements.  In practice, this difficulty is usually circumvented by holding most of the nucleons fixed and only allowing a few to occupy single particle states above the lowest energy.  While this approach works reasonably well for the lowest-lying nuclear states, it's efficacy breaks down at higher energies (higher nuclear energies imply more nucleons above the lowest single particle energies) and when the model has too few single particle states, {\it i.e.} is restricted, allowing too few basis states to yield a realistic set of total nuclear eigenstates.

Because of these computational obstacles and the fact that we want to understand the GT structure of {\it very} highly excited nuclei, we are relegated in this work to studying relatively light nuclei.  The biggest drawback of this approach is that although light nuclei are abundant prior to the onset of collapse, they are disfavored during infall.  In our favor, reference \cite{mbf:2013} found that in some respects, heavy nuclei and light nuclei exhibit similar weak transition characteristics.  In any case, light nuclei are at present the only option for computing highly excited states, and we will ideally learn something that will shed light on the behavior of all nuclei, including heavier, more neutron rich species.

In section \ref{sec:sm_gt}, we provide a brief overview of the nuclear shell model and GT transitions, as it will be convenient in later sections to have that picture in mind.  Section \ref{sec:previous} outlines the historical approach to the problem at hand and discusses its weaknesses.  The results of our electron capture strength computations are in section \ref{sec:results}, and using those results, we show calculations of electron capture rates in section \ref{sec:rate}.  We give discussion and conclusions in section \ref{sec:discussion}.

\section{Nuclear Shell Model and GT Transitions}
\label{sec:sm_gt}
In the shell model, individual nucleons are considered to occupy non-interacting single-particle states, with the sets of occupied states (configurations) coupled to have good spin J and isospin T.  Energy, angular momentum, and isospin eigenstates can be constructed by diagonalizing a residual nucleon-nucleon Hamiltonian in the configuration basis.  This mixes many configurations into a single nuclear state:
\begin{equation}
\vert\Psi_{J,T}\rangle_i=\sum_{k} A_{ik}\vert C_{J,T}\rangle_k
\end{equation}
where $\vert\Psi_{J,T}\rangle_i$ is nuclear eigenstate $i$ with spin J and isospin T, the A$_{ik}$ are complex amplitudes, and $\vert C_{J,T}\rangle_k$ is the $k$th configuration with spin J and isospin T.

One-body nuclear transitions -- such as the Gamow-Teller transition -- consist of a single nucleon changing its single particle state.  There are three qualitatively different single particle GT transitions: spin flip transitions (from an $l+\frac{1}{2}$ state to an $l-\frac{1}{2}$ state), back spin flip transitions (from $l-\frac{1}{2}$ to $l+\frac{1}{2}$), and lateral transitions (no change in total angular momentum).  Respectively, these represent a net gain, loss, and no change in single particle energy up to differences in energy between neutron and proton single particle states.  If a nuclear state has as one of its components a configuration resulting from a single particle transition in a particular initial state, then the nucleus can transition to that final state.  The strength of the transition from an initial nuclear state $\vert i\rangle$ to a final state $\vert f\rangle$ is given by
\begin{equation}
\vert\langle f\vert\sum_k\widehat{o}_k\vert i\rangle\vert^2
\end{equation}
where $\widehat{o}_k$ is a single body operator on the k$^{th}$ nucleon.  Throughout this paper, ``GT strength'' or ``electron capture strength'' will refer to the reduced nuclear transition probability B(GT)$_{if}$, given by
\begin{equation}
\frac{\vert\langle f\vert\vert\Sigma_k(\vec{\sigma}\tau_{-})_k\vert\vert i\rangle\vert^2}{2J_i+1}
\end{equation}
where $\vec{\sigma}\tau_{-}$ is the one-body Gamow-Teller lowering operator and the sum is over nucleons.

\section{Previous Adaptation of GT Brink-Axel Hypothesis}
\label{sec:previous}
FFNII \cite{ffn:1982a} approached the problem of GT transition strength distributions by using experimental values of the strength where known, supplementing that with estimated allowed and forbidden strength to known states in the daughter nucleus, and placing the remainder of the GT strength computed from a zero-order shell model into a single narrow resonance at an energy also computed using a zero-order shell model.  Using two simple assumptions, FFNII took the strength and relative energy of the resonance to be the same for all excited states as it is for the ground state.  First, assume that the individual nucleons are distributed among the single particle states in a way that is {\it on average} independent of nuclear excitation energy.  Second, assume that the transition energy of the GT resonance is principally due to a single nucleon undergoing a spin flip, and thus is similar in excited states to that of the ground state.  To the extent that these approximations are valid, they are extremely useful, as the partition function becomes algebraically irrelevant in determining the resonant electron capture rate.  From FFNII, the total electron capture rate through resonant transitions is given by
\begin{equation}
\Lambda^{res} = \sum\limits_{i} P_{i}\lambda_{i}^{res}
\end{equation}
where P$_i$ is the probability that the nucleus is in state $\vert i\rangle$ (given by the product of the degeneracy and the Boltzmann factor, divided by the partition function) and $\lambda_{if}^{res}$ is the resonant transition rate from state $\vert i\rangle$ to state $\vert f\rangle$, itself a function of nuclear structure and electron distribution in the supernova core.  But under the assumption that the GT resonances are the same--irrespective of nuclear excitation energy--all of the $\lambda_{i}^{res}$ are identical; we shall call them $\lambda^{res}$.  We now have
\begin{eqnarray}
\Lambda^{res}&=&\sum\limits_{i}P_{i}\lambda^{res} \nonumber \\
&=&\lambda^{res}\sum\limits_{i}P_{i} \nonumber \\
&=&\lambda^{res}
\end{eqnarray}

So, the total resonant transition rate is simply the resonant transition rate of any single state, which we take to be the ground state.  Of course, highly excited states in the parent would be in the GT resonances of lower energy states in the daughter, leading to ``back-resonant" transitions.  Accounting for the fact that the P$_i$ will not sum to unity for back-resonant transitions (low-lying initial states have no back-resonance) and otherwise treating them identically to resonant transitions, we eventually arrive at
\begin{equation}
\Lambda^{back res}=\lambda^{back res}\frac{G^{d}}{G^{p}}e^{\frac{-R}{kT}}
\end{equation}
where G$^p$ (G$^d$) is the partition function of the parent (daughter) nucleus and R is the characteristic transition energy of the GT resonance from the daughter nucleus to the parent.  Finally, Fermi transitions are handled in an identical manner to the GT transitions, and the rates are summed along with the rates from known and estimated transitions to get the total capture rate.

A priori, we might expect the GT Brink-Axel hypothesis to fail.  If we keep the assumption that single particles are distributed roughly independently of nuclear excitation energy, we should be unsurprised if the GT resonance moves dramatically or is redistributed in transition energy, since at sufficiently high initial excitation, there will be strength for the daughter nucleus to be at {\it many} energies relative to the parent, without any particular single particle transition dominating the strength.  By assumption the single particles in all of these daughter states are also arranged similarly, so we would rather expect the GT strength to be broadly distributed in transition energy.  The question, then, is in what way does the hypothesis fail?  Do strength distributions evolve in some characteristic way as initial excitation energy increases, or must we abandon the hypothesis completely and replace it with a thermal mean strength distribution?

\section{GT Strength Computations}
\label{sec:results}
Using the shell model code Oxbash \cite{oxbash}, we performed shell model calculations of $A=28$ nuclei using a closed $^{16}$O core and 12 valence nucleons in the sd shell.  Although $A=28$ is unrealistically light for the supernova core environment,  we chose to use it because it provides a good balance of complexity and computability; that is, we have many valence nucleons and holes (implying many single particle configurations), but there are few enough configurations that we can compute nuclear eigenstates in a reasonable time.  We also performed a computation of $^{24}$Mg using the same interaction, but with 8 valence nucleons.

The sd shell consists of the single particle sates 0d$_{5/2}$, 1s$_{1/2}$, and 0d$_{3/2}$.  In these computations, we used the USDB Hamiltonian \cite{br:2006}, with single particle energies $-3.9257$, $-3.2079$, and $2.1117$ MeV, respectively.  In the GT interaction, nucleons can transition from 2s$_{1/2}$ to 2s$_{1/2}$, and from either d sub-orbital to either d sub-orbital.  

In order to make a comparison with the FFN results, we need to address quenching \cite{brownwildenthal:1985}.  FFN implemented quenching as follows: (1) experimentally-determined and ``guessed'' relatively low-lying vector and axial vector transitions are, of course, already ``quenched''; (2) calculated Gamow-Teller resonance transitions were deliberately not quenched.  We follow the same procedure here to facilitate comparison with FFN.  As detailed in FFNIV, inspection of the effective log-ft values indicate if stellar weak rates are dominated by low-lying transitions (regime 1) or resonance transitions (regime 2).  Since we are mostly concerned here with high density and temperature, \emph{usually} the stellar weak rates are resonance-dominated, \emph{i.e.} in regime 2.  Here we do not quench our calculated rates where unmeasured, calculated GT strength is involved \-- again, simply to facilitate comparison with FFN.  However, we recommend quenching wherever sd and fp shell model strength is used to compute rates for astrophysical or any other use.

\subsection{$^{28}$Si}
\begin{figure}
\centering
\includegraphics[scale=0.39]{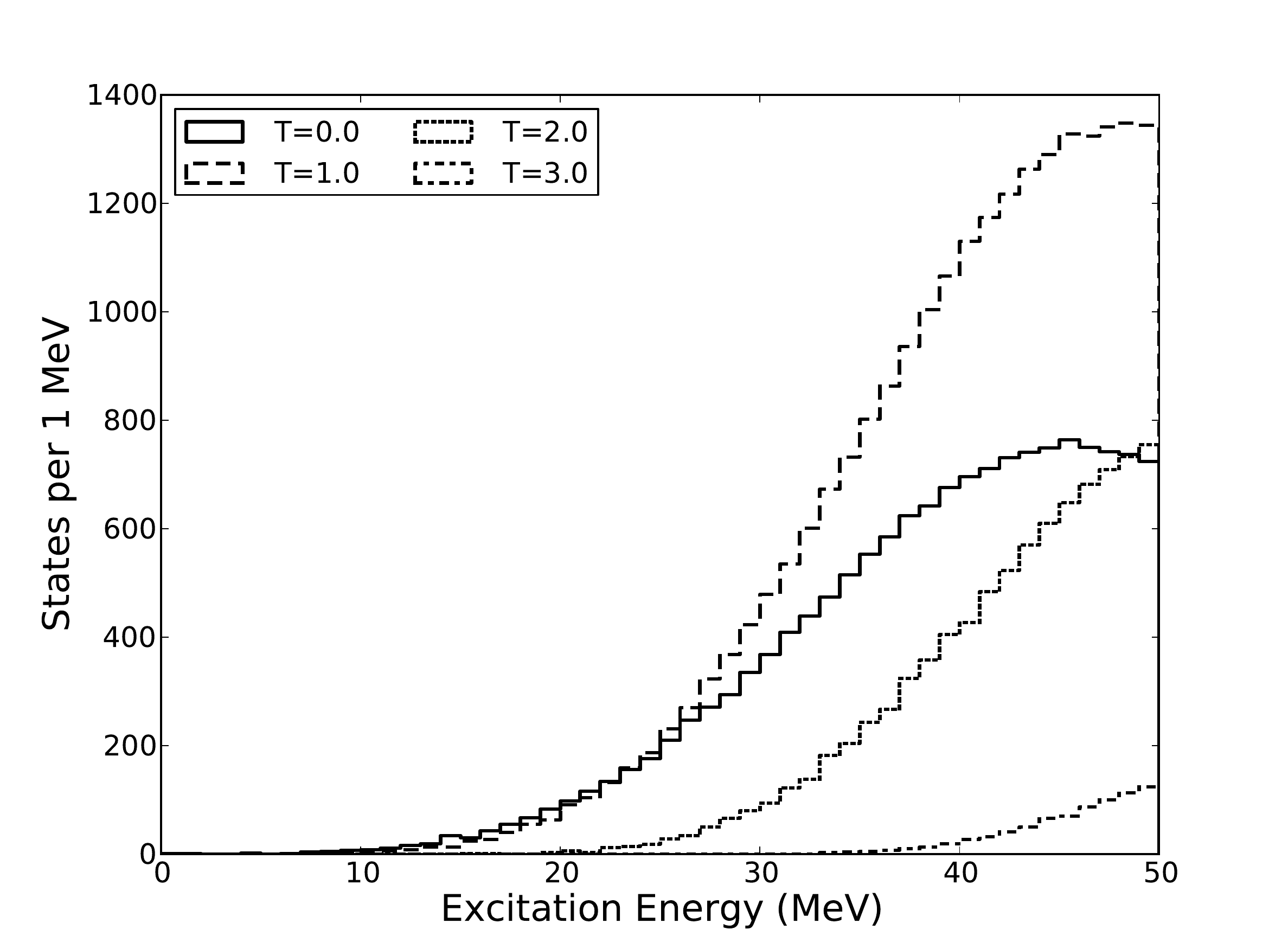}
\caption{$^{28}$Si density of states per 1 MeV as functions of isospin.  The inflection points for $T=0$ and $T=1$ near 30 MeV--which indicate a marked departure from the expected exponential growth--suggest we should be wary of results for states with energies above the mid-20s.}
\label{fig:si_dos}
\end{figure}

We first examined $^{28}$Si.  Although this nucleus is neutron poor by supernova collapse standards, it has the most single particle configurations among sd-shell nuclei and therefore computationally is the most realistic.  Figure \ref{fig:si_dos} shows the density of states per 1 MeV for this nucleus broken down by isospin.  The inflection points mark the regions where the density of states departs radically from an exponential form, indicating a departure from our expectation for reality, in turn implying that results for states with energies near and above the inflection point may be significantly impacted by the model space restriction.  The inflection points on the $T=0$ and $T=1$ curves occur a little above 30 MeV, so we will treat states with energies above the mid-20s of MeV with circumspection.

\begin{figure}
\centering
\includegraphics[scale=.33]{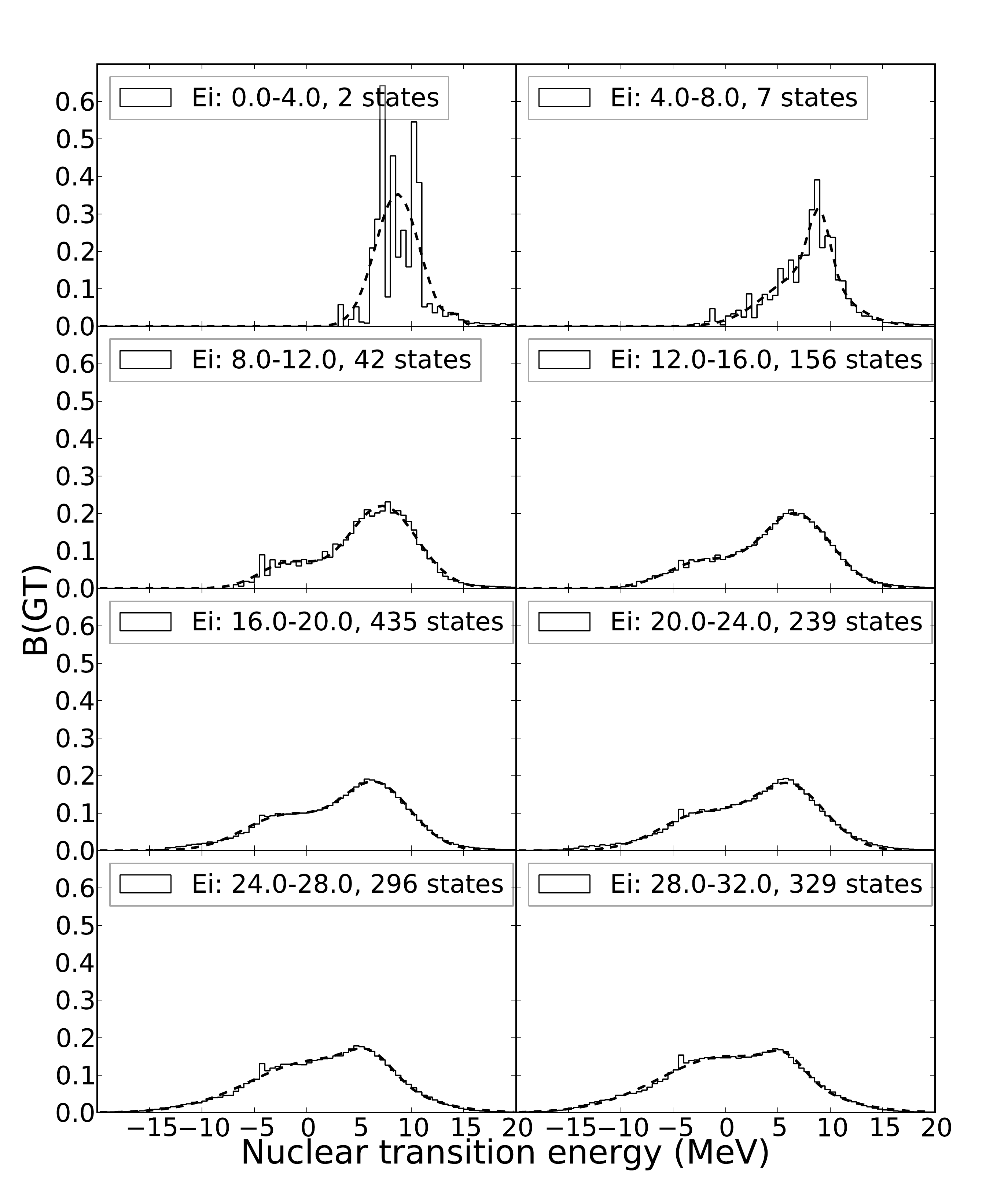}
\caption{Average Gamow-Teller strength distribution in $^{28}$Si as a function of initial excitation energy E$_i$.  Strength is binned in 0.5 MeV increments of transition energy.  Also shown are fits of the distributions to the double Gaussian in equation \ref{eq:dbl_gauss}.  The dependence on initial excitation energy becomes small at high excitation.}
\label{fig:si_str}
\end{figure}

Figure \ref{fig:si_str} shows the electron capture strength distribution in 0.5 MeV transition energy bins as a function of excitation energy and nuclear transition energy (that is, the total energy input required make the transition, including the change in nuclear mass); the distributions are averaged over the indicated number of states in each parent nucleus excitation energy bin.  We found that the GT Brink-Axel hypothesis as originally formulated does not obtain in that the strength distributions of excited states bear no resemblance to the ground state.  However, the GT strength distribution is almost independent of initial state energy for transitions proceeding from initial excitations greater than 12 or 16 MeV.  There appears to be some energy dependence above 24 MeV excitation, though this may be due to the limitations of the model space.  Figure \ref{fig:si_str} also shows fits of the strength distributions to a double Gaussian of the form
\begin{equation}
C_1e^{-(\Delta E-\Delta E_1)^2/2\sigma_1^2}+C_2e^{-(\Delta E-\Delta E_2)^2/2\sigma_2^2}.
\label{eq:dbl_gauss}
\end{equation}
The fit parameters for strength density are shown in table \ref{table:si_params}.  Note that the computed strength distributions in figure \ref{fig:si_str} are histograms, so the fit curves are scaled vertically to account for the effect of the particular choice of transition energy bin width.  This analysis confirms that the distributions are weak functions of parent nucleus energy at high excitation.

\begin{table}
\begin{tabular}{ | c | c | c | c | c | c | c | c | c | }
\hline
E$_0$ & C$_1$ & $\Delta$E$_1$ & $\sigma_1$ & B(GT)$_1$ & C$_2$ & $\Delta$E$_2$ & $\sigma_2$ & B(GT)$_2$ \\
\hline
0 - 4 & -- & -- & -- & -- & 0.70 & 8.7 & 2.1 & 3.7 \\
4 - 8 & 0.28 & 7.5 & 3.6 & 2.5 & 0.38 & 8.9 & 1.1 & 1.0 \\
8 - 12 & 0.13 & -1.8 & 2.6 & 0.85 & 0.44 & 7.2 & 3.3 & 3.6 \\
12 - 16 & 0.14 & -2.3 & 3.3 & 1.2 & 0.40 & 6.7 & 3.3 & 3.3 \\
16 - 20 & 0.18 & -2.2 & 3.8 & 1.7 & 0.36 & 6.6 & 3.3 & 3.0 \\
20 - 24 & 0.19 & -2.3 & 3.9 & 1.9 & 0.34 & 6.1 & 3.4 & 2.9 \\
24 - 28 & 0.28 & 1.1 & 6.4 & 4.5 & 0.13 & 6.2 & 2.2 & 0.72 \\
28 - 32 & 0.30 & 0.24 & 6.5 & 4.9 & 0.11 & 5.8 & 1.8 & 0.50 \\
\hline
\end{tabular}
\caption{Double Gaussian fit parameters and total strength of each peak for $^{28}$Si.  E$_0$ is the initial excitation energy in MeV, and the parameters are as shown in equation \ref{eq:dbl_gauss}.  The C$_i$ are dimensionless, and the other parameters are in units of MeV.}
\label{table:si_params}
\end{table}

\begin{figure}
\centering
\includegraphics[scale=0.6]{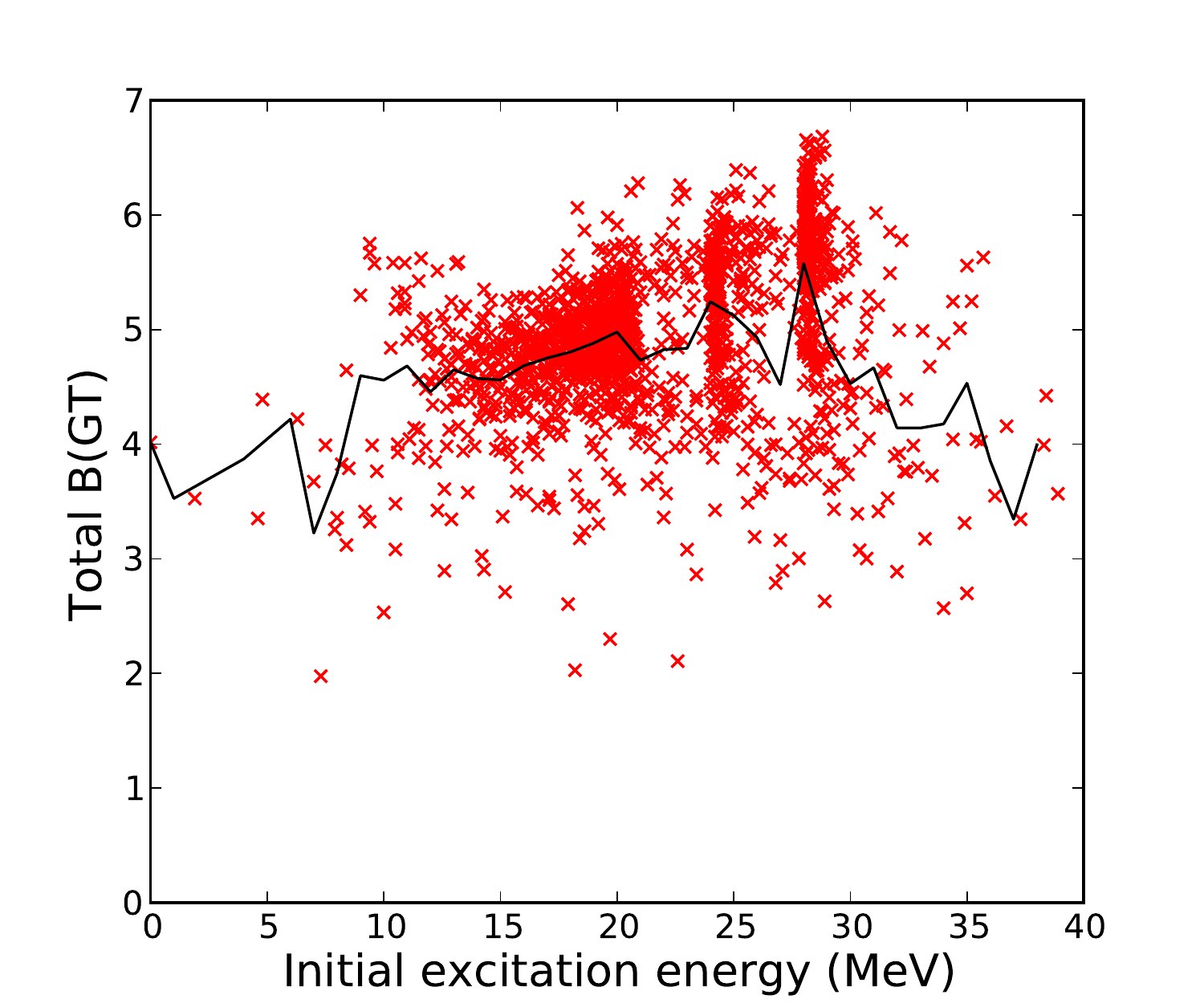}
\caption{Total GT strength in $^{28}$Si as a function of excitation energy.  Each point corresponds to an individual state computed from the shell model, giving a GT sum rule for that state.  The black line shows the average total strength in 1 MeV bins.}
\label{fig:si_tot_str}
\end{figure}

Figure \ref{fig:si_tot_str} shows the total GT strength vs. excitation energy for our shell model states, with each point corresponding to a single initial state.  The vertical stripes are due to sampling; all shell model states up to 20 MeV are included, as are many states near 24 and 28 MeV.  The black line shows the total strength for all included states averaged over 1 MeV bins.  Where the sampling is dense, there is an overall positive trend in total strength with excitation energy.

\begin{figure}
\centering
\includegraphics[scale=.6]{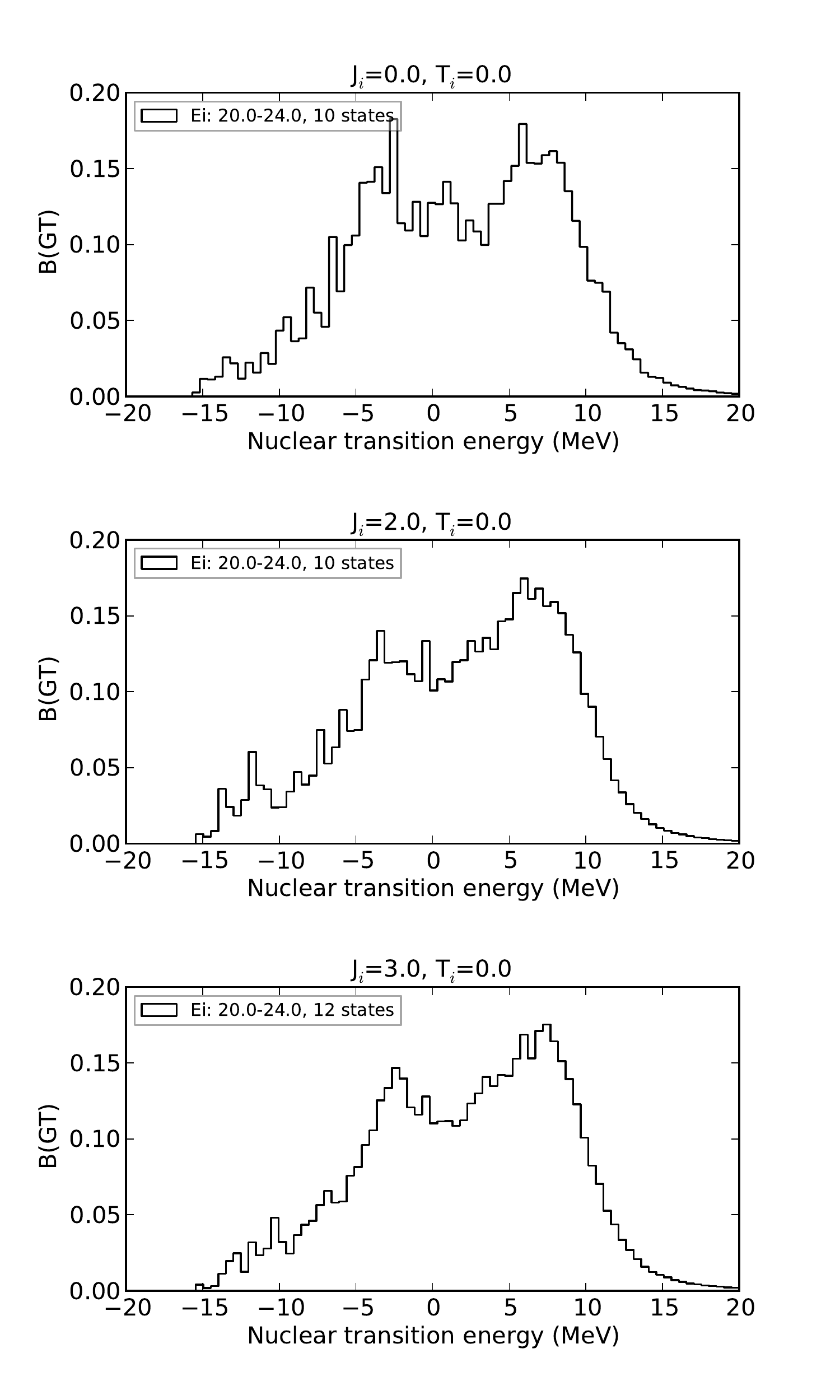}
\caption{Gamow-Teller strength distribution in $^{28}$Si at initial excitation energy $E_i=20-24$ MeV with initial isospin $T_i=0$ as a function of initial spin $J_i$.  The strength distribution is not strongly dependent on $J_i$.  We saw this trend in all nuclei we studied.}
\label{fig:si_str_j}
\end{figure}

Decomposing the strength distributions into contributions from states with specific initial spin and isospin reveals that the trend of figure \ref{fig:si_str} holds; that is, regardless of choice of a particular initial spin and/or isospin, the Brink hypothesis fails at low excitation, but is recovered at high excitation.  Furthermore, nuclear spin is not an important contributor in determining either the shape or total strength of the distribution.  Figure \ref{fig:si_str_j} shows the distributions for a representative selection of spins with initial isospin $T_i=0$ in the $E_i=20-24$ MeV bin.  We observed this pattern of $J_i$-independence at all excitation energies and isospins.

\begin{figure}
\centering
\includegraphics[scale=.6]{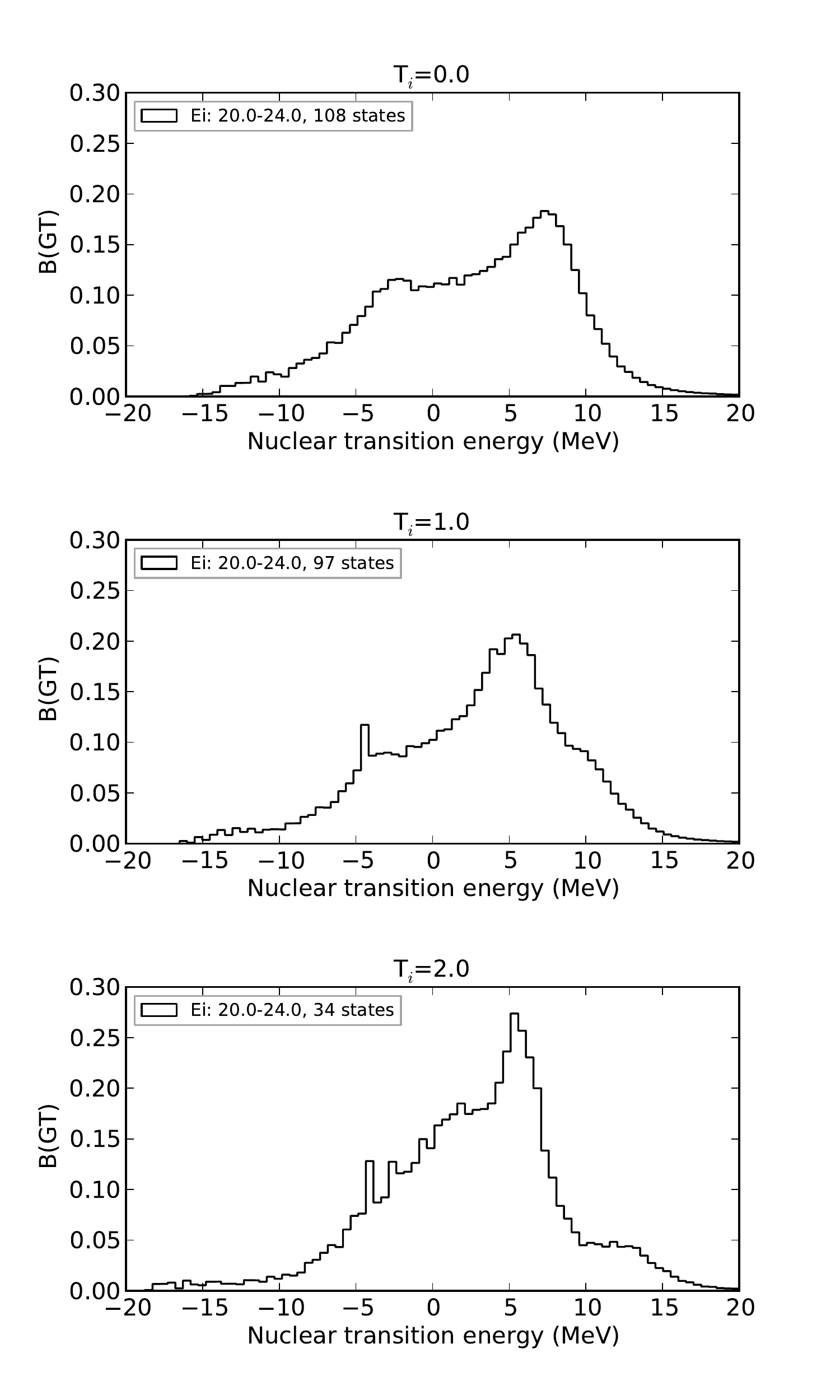}
\caption{Gamow-Teller strength distribution in $^{28}$Si at initial excitation energy $E_i=20-24$ as a function of initial isospin $T_i$.  The strength distribution has an apparent dependence on initial isospin.}
\label{fig:si_str_t}
\end{figure}

Turning our attention now to isospin, we find that isospin \emph{does} play a role in the distribution of GT strength.  In figure \ref{fig:si_str_t} we show strengths for the $E_i=20-24$ MeV bin.  Each panel gives the distribution for a different $T_i$, but because nuclear spin does not strongly affect the distribution, we include in figure \ref{fig:si_str_t} \emph{all} values of $J_i$.  As isospin increases, the locations and strengths of the peaks and plateaus in the distribution shift.

\begin{figure}
\centering
\includegraphics[scale=.6]{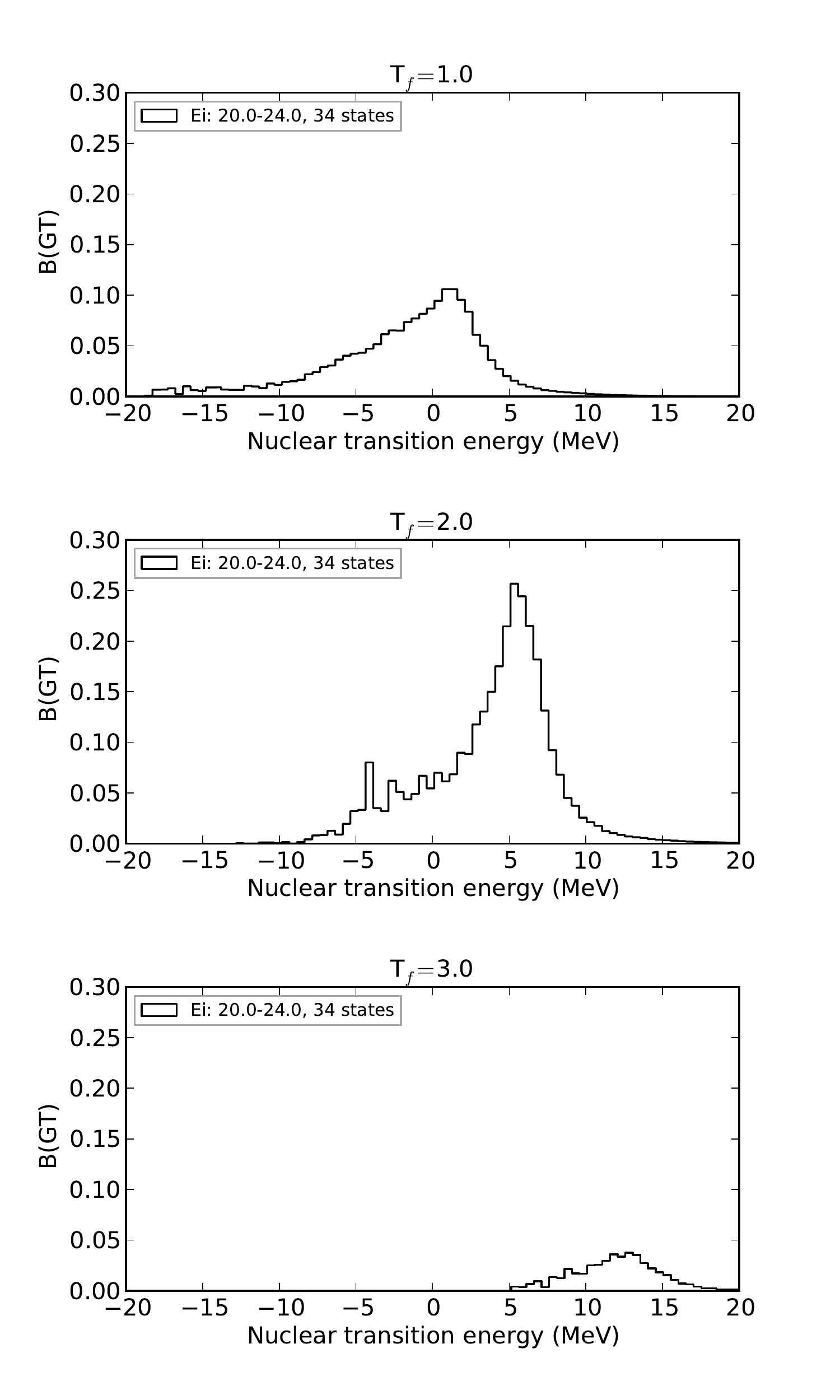}
\caption{Strength distribution for $^{28}$Si with initial isospin $T_i=2$ as a function of final state isospin.  Comparison with figure \ref{fig:si_str_t} shows that certain features of the total strength distribution are consequences of the distributions to specific final state isospins.}
\label{fig:si_str_dec}
\end{figure}

The shapes of the single initial isospin strength distributions can be partially understood by decomposing them into contributions from final states with specific isospin.  Although the level density is dominated by $T=0$ and $T=1$ states in the energy range of interest for the supernova problem, we examine here $T_i=2$ because it has a greater number of final isospins and therefore is more illustrative of the effect.  Figure \ref{fig:si_str_dec} shows the $T_f$-decomposition for $T_i=2$ states in the $E_i=20-24$ MeV bin.  Evidently, the strength distribution is strongly dependent on the final isospin with distinct consequences on the shape of the full distribution.  For example, the large peak in the total strength distribution at $\Delta E\approx 5$ MeV for states with $T_i=2$ is due to transitions to final states with $T_f=2$, and the small peak at $\Delta E\approx 13$ MeV is due to transitions to final states with $T_f=3$.

\begin{figure}
\centering
\includegraphics[scale=.6]{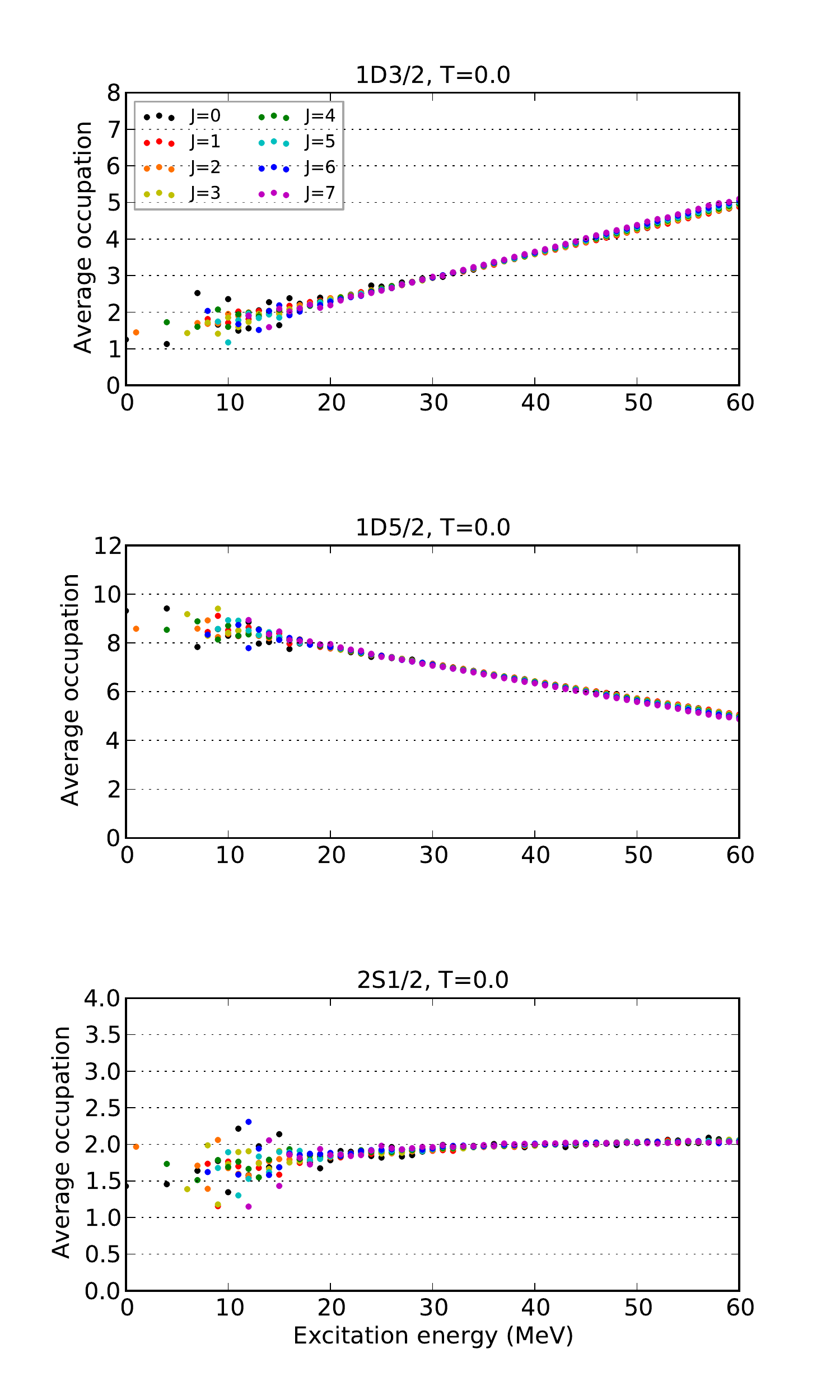}
\caption{$^{28}$Si single particle state occupation for nuclear states with isospin $T=0$.  The occupation numbers are the sum of protons and neutrons.  The linear dependence on excitation energy of the d orbital occupation numbers is understood to arise from the spin-orbit splitting and particle-hole repulsion energies of those orbitals.  This dependence is consistent across all values of T, although the intercepts of the d orbitals do shift as T increases.}
\label{fig:si_sps_occ}
\end{figure}

Finally, we sought an understanding of the similarity of the strength distributions in the high excitation energy regime.  To this end, we examined single particle distribution as a function of nuclear excitation energy.  Figure \ref{fig:si_sps_occ} shows the average single particle state occupations as functions of spin and excitation energy for $T=0$ states in $^{28}$Si.  The most salient features are that the occupations have no clear dependence on nuclear spin, and the 1d state occupations have a linear dependence on nuclear excitation with slopes of roughly 1 particle per 12 MeV (which is approximately the spin-orbit splitting energy $+$ particle-hole repulsion energy in this sub-shell), while the 2s$_{1/2}$ occupation is independent of excitation; this is in contrast to FFNII \cite{ffn:1982a}, which assumed that the average occupations of all single particle states were independent of nuclear excitation energy.  While figure \ref{fig:si_sps_occ} shows only T$=0$ states, the trends are consistent for all isospins, with the exception that the intercept of the 1d$_{3/2}$ (1d$_{5/2}$) occupation gradually shifts by -1 (1) particle as T goes from 0 to 3, and shifts an additional -1 (1) particle as T goes from 3 to 4, but this shift may be due to model space restrictions.

Since we did the computations in this paper with isospin as a good quantum number, we can take the single particle occupations in figure \ref{fig:si_sps_occ} to be split proportionately between the valence protons and neutrons.  In the case of $^{28}$Si, then, the proton and neutron single particle occupation numbers are each 1/2 the total occupation.  This implies that the 1 particle per 12 MeV slope in fig. \ref{fig:si_sps_occ} is split evenly between protons and neutrons, giving a slope for each species of 1 particle per 24 MeV.  Perhaps, then, the assumption in \cite{ffn:1982a} that the single particle distributions are all similar to the ground state can be simply revised to say that above a certain nuclear excitation energy, the single particle distributions change only very slowly with excitation energy, resulting ultimately in similarly slowly changing strength distributions.  This leaves us to challenge the second assumption in that work: that the transition energy of the GT resonance does not change with nuclear excitation energy.

The spin flip single particle transition dominates in electron capture from the ground state, resulting in the resonance at the observed energy.  However, at higher excitation energies there is an abundance of final nuclear states that are reachable by the other single particle transitions that can leave the daughter nucleus at similar or lower excitation.  Thus, the GT strength distribution changes with increasing initial excitation, spreading to lower transition energy.  Comparing the relative positions of the peaks in the strength distributions of all nuclei considered in this paper suggests there may be a correlation between the single particle state and particle-hole repulsion energies with the locations of the peaks in the strength distributions, but we will not further explore that in this paper.

Ultimately, given that single particle state occupations vary slowly at high excitation energy, it is unsurprising that over a broad range of energy (above the rapid variation at low energy and below where the density of states falls below exponential growth), the strength distributions are largely independent of excitation.

\subsection{$^{24}$Mg}
\begin{figure}[here]
\centering
\includegraphics[scale=0.6]{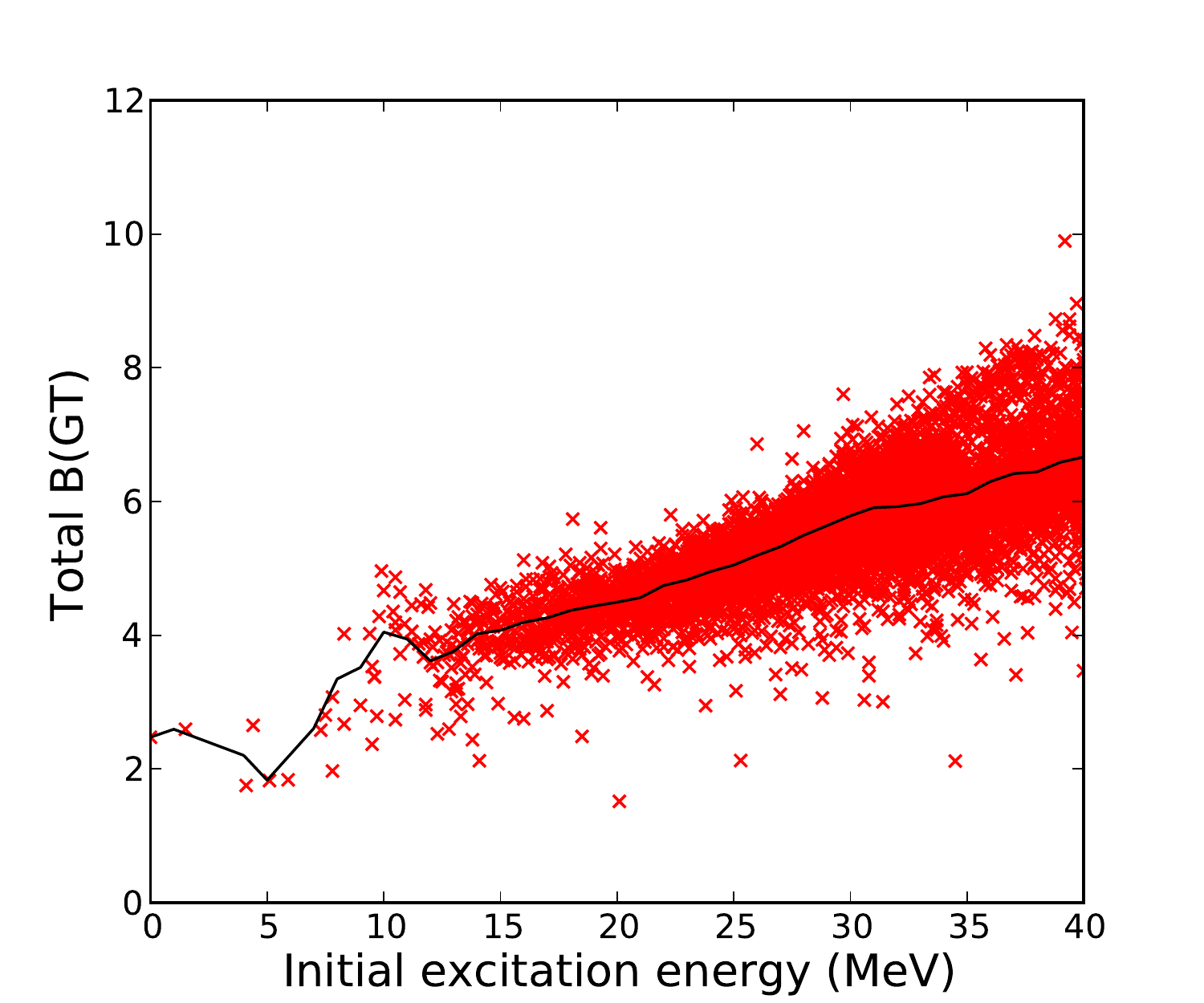}
\caption{Total GT stregth in $^{24}$Mg as a function of excitation energy.  The black line shows the average total strength, computed from 1 MeV bins.}
\label{fig:mg_tot_str}
\end{figure}

For the sake of connecting our $^{28}$Si results with earlier work, we computed the Gamow-Teller strengths for $^{24}$Mg.  Frazier et al \cite{fbmz:1997} examined total strength as a function of excitation and found results similar to ours, given in figure \ref{fig:mg_tot_str}.  The gradual increase of total strength with initial excitation corroborates the result for $^{28}$Si (figure \ref{fig:si_tot_str}) and suggests it is a general feature of nuclei, at least in the sd shell.

\begin{figure}
\centering
\includegraphics[scale=0.39]{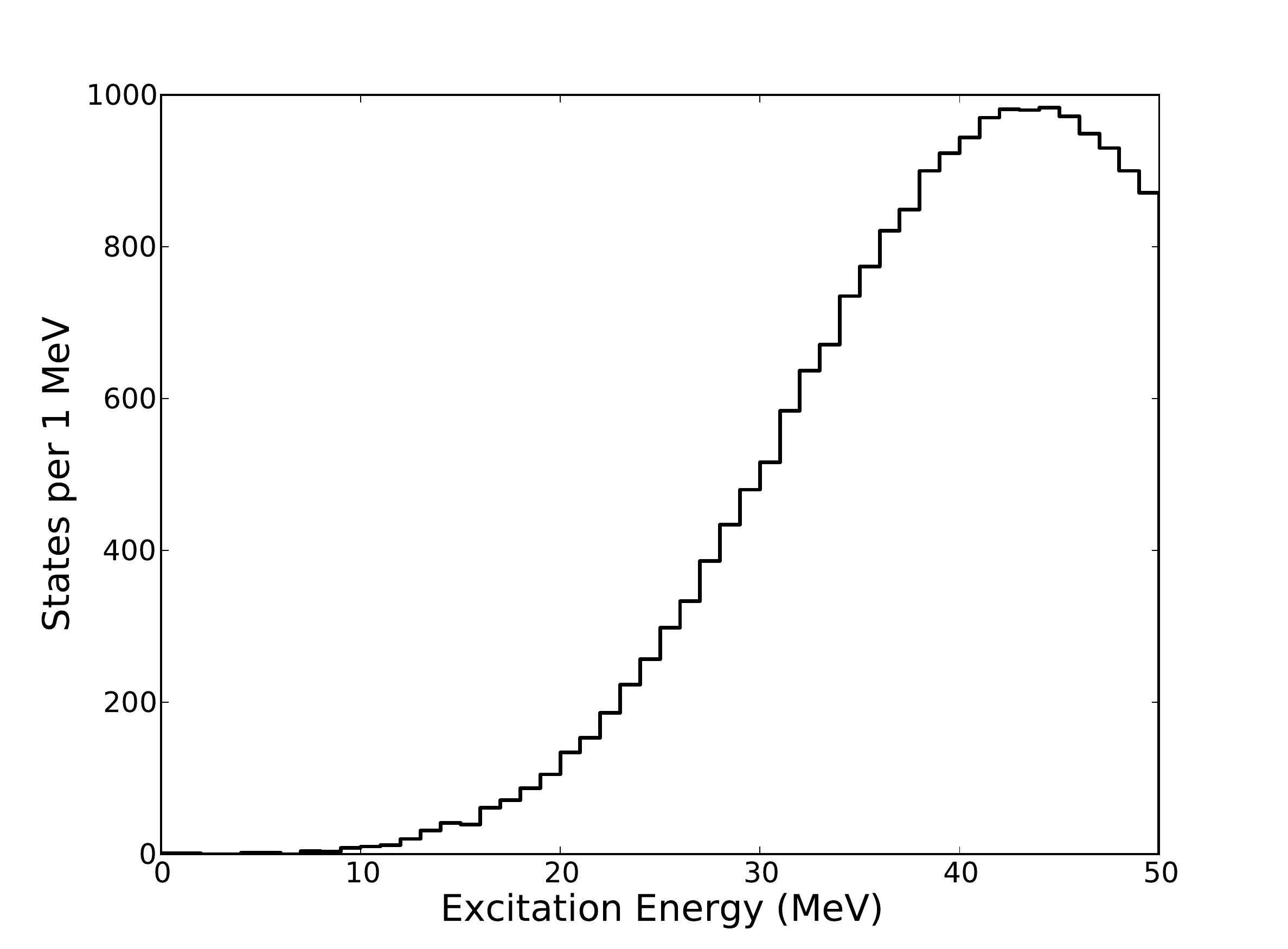}
\caption{$^{24}$Mg density of states.  The inflection point is near 28 MeV, indicating that we cannot be confident of strength distributions for initial state energies above the low 20s.}
\label{fig:mg_dos}
\end{figure}

\begin{figure}
\centering
\includegraphics[scale=0.33]{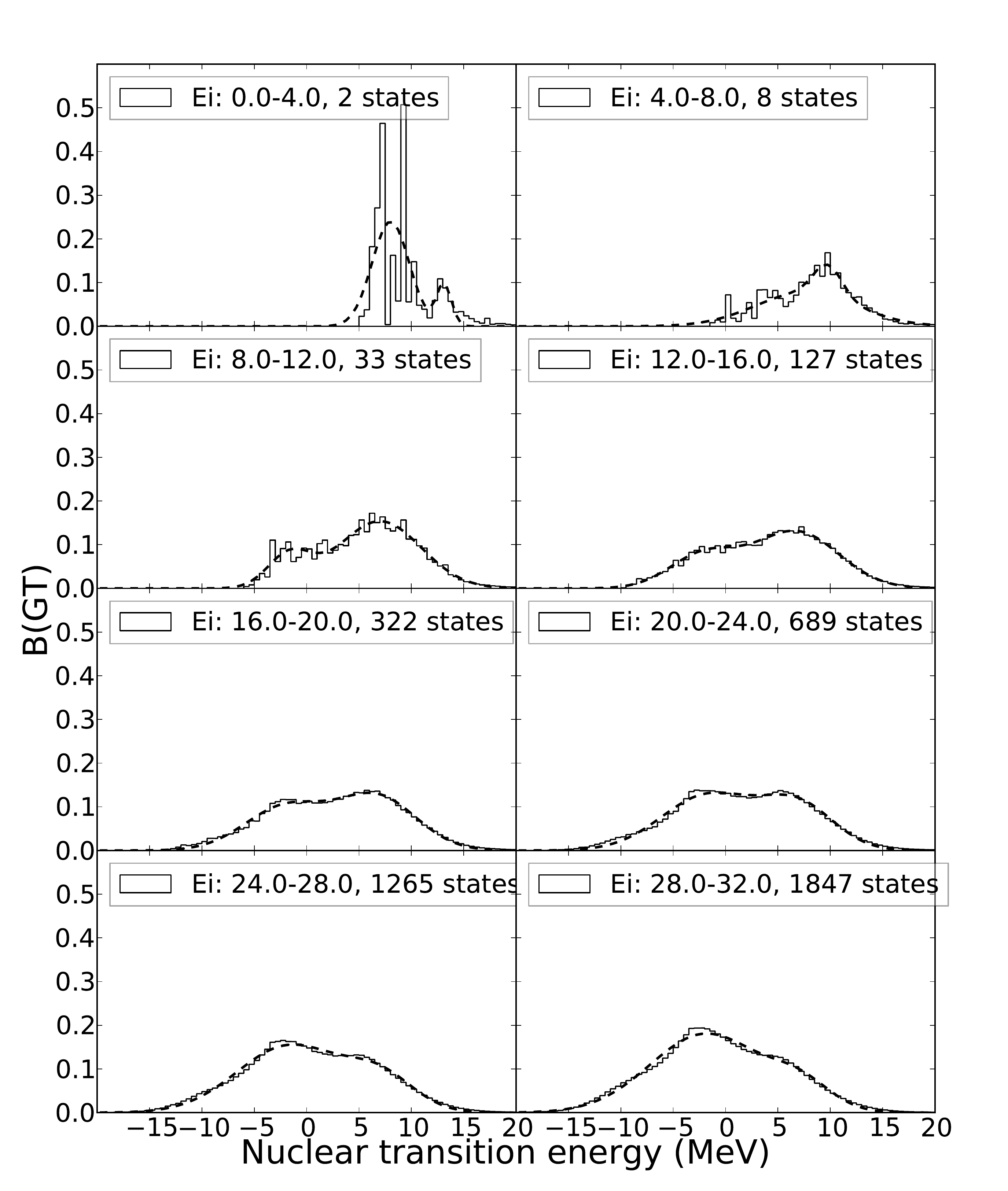}
\caption{Computed average Gamow-Teller strength distribution in 0.5 MeV transition energy bins and fits to a double Gaussian in $^{24}$Mg as a function of initial excitation energy $E_i$.  As in the analysis of $^{28}$Si, the fit curves are scaled to account for the choice of transition energy bin width.  Between the $E_i=12-16$ MeV and $20-24$ MeV bins, the strength varies slowly with excitation.}
\label{fig:mg_str}
\end{figure}

The density of states for $^{24}$Mg (figure \ref{fig:mg_dos}) has an inflection point near 28 MeV, so strength distributions for states with initial energies above the low 20s are suspect.  The strength distributions for $^{24}$Mg behave qualitatively the same as $^{28}$Si.  That is, above 12 MeV (and below the model space restriction range), the distributions (figure \ref{fig:mg_str}) are not strong functions of initial excitation.  Table \ref{table:mg_params} shows that where the distribution is stable (between 12 and 20 MeV excitation), the fit parameters agree with those for $^{28}$Si.  This bodes well for extrapolating to other nuclei.

\begin{table}[here]
\begin{tabular}{ | c | c | c | c | c | c | c | c | c | }
\hline
E$_0$ & C$_1$ & $\Delta$E$_1$ & $\sigma_1$ & B(GT)$_1$ & C$_2$ & $\Delta$E$_2$ & $\sigma_2$ & B(GT)$_2$ \\
\hline
0 - 4 & 0.48 & 8.0 & 1.8 & 2.2 & 0.19 & 13.1 & 0.70 & 0.33 \\
4 - 8 & 0.16 & 7.7 & 4.8 & 1.9 & 0.14 & 9.8 & 1.3 & 0.46 \\
8 - 12 & 0.15 & -1.7 & 2.1 & 0.79 & 0.30 & 7.0 & 4.0 & 3.0 \\
12 - 16 & 0.15 & -2.1 & 3.3 & 1.2 & 0.26 & 6.7 & 4.1 & 2.7 \\
16 - 20 & 0.20 & -2.0 & 3.9 & 2.0 & 0.24 & 6.7 & 3.8 & 2.3 \\
20 - 24 & 0.26 & -1.6 & 4.5 & 2.9 & 0.20 & 7.0 & 3.5 & 1.8 \\
24 - 28 & 0.30 & -1.6 & 5.0 & 3.8 & 0.15 & 7.1 & 3.3 & 1.2 \\
28 - 32 & 0.36 & -1.9 & 5.4 & 4.9 & 0.11 & 7.2 & 2.9 & 0.80 \\
\hline
\end{tabular}
\caption{Double Gaussian fit parameters and total strength of each peak for $^{24}$Mg.  The good agreement with $^{28}$Si means extrapolation to other nuclei may be tenable.}
\label{table:mg_params}
\end{table}

\subsection{$^{28}$Mg}
\begin{figure}
\centering
\includegraphics[scale=0.39]{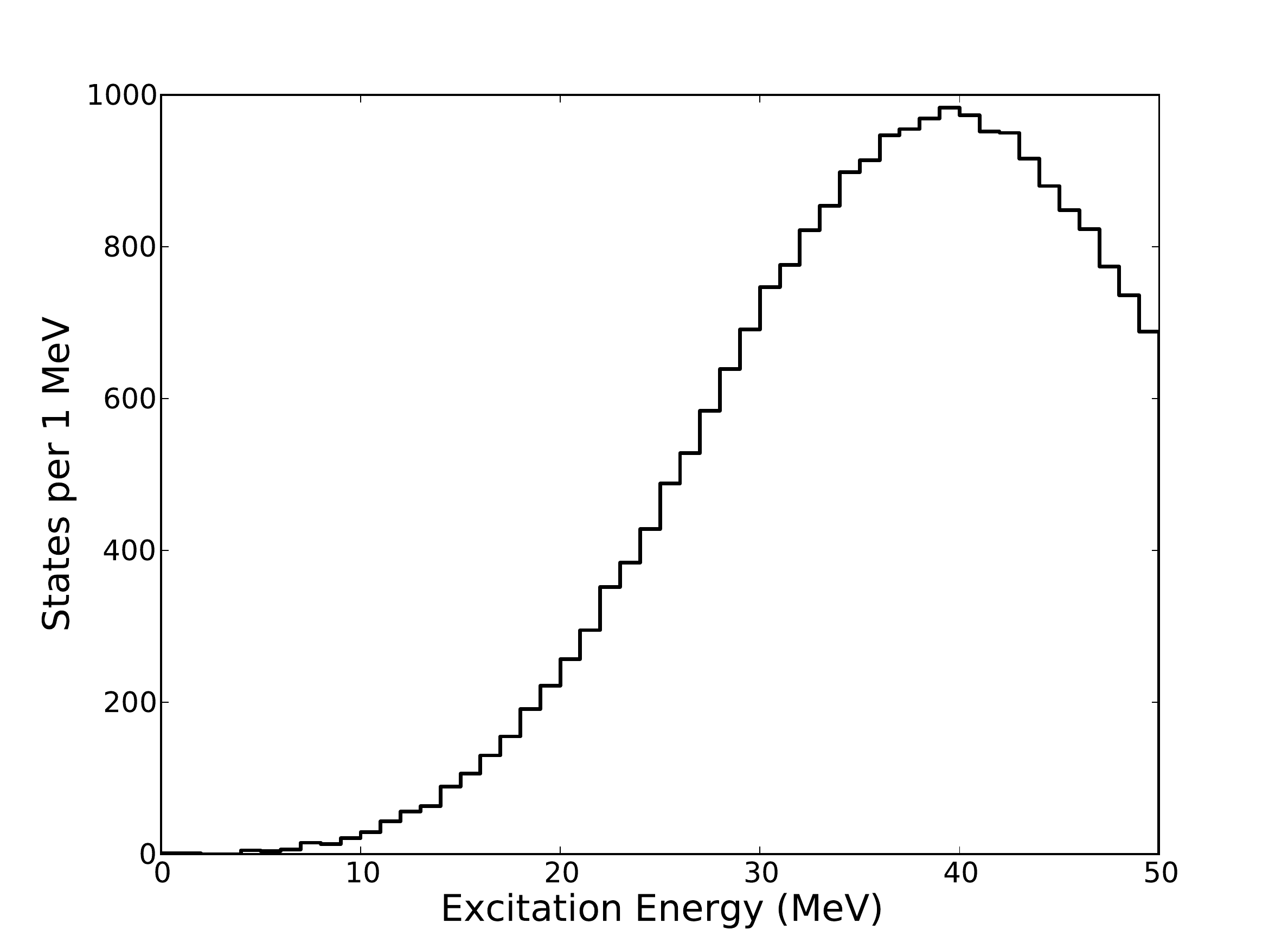}
\caption{$^{28}$Mg density of states.  The inflection point at $\sim$22 MeV predicts that the region where the strength distribution is stable will be narrow.}
\label{fig:mg28_dos}
\end{figure}

\begin{figure}
\centering
\includegraphics[scale=0.33]{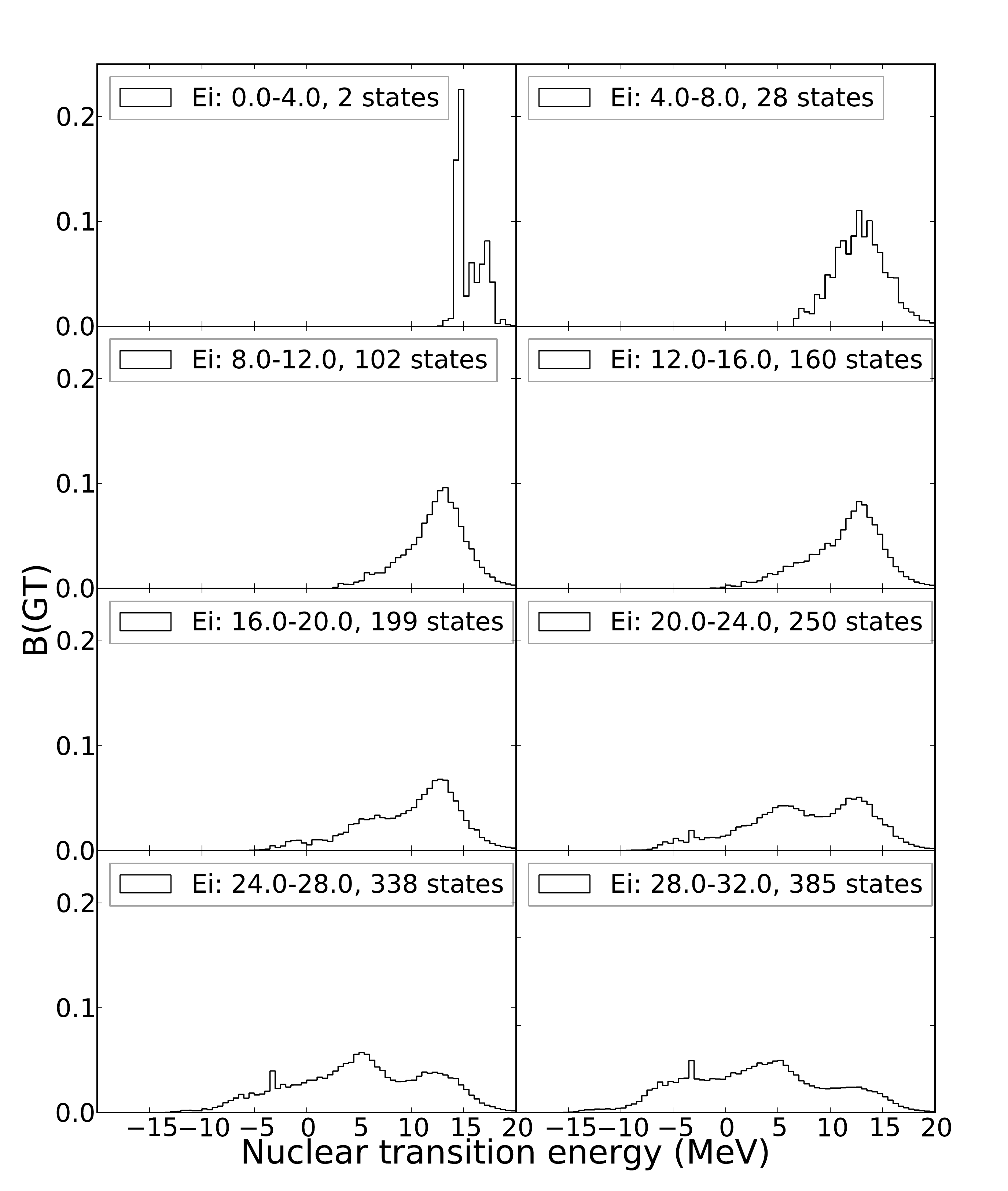}
\caption{$^{28}$Mg strength distribution.  The distribution stabilizes briefly between 8 and 16 MeV before model space restriction impacts the results.}
\label{fig:mg28_str}
\end{figure}

Figure \ref{fig:mg28_dos} shows the density of states for $^{28}$Mg; the inflection point occurs at $E\approx 22$ MeV.  The low energy of the inflection really crowds the region where we expect the strength distribution to be stable, and this indeed manifests out in figure \ref{fig:mg28_str}.  The high transition energy peak in the distribution appears to stabilize briefly around 8-16 MeV, but it rapidly falls off and the low transition strength grows as the initial excitation goes into the model space restricted region.

\subsection{$^{28}$Na}
\begin{figure}
\centering
\includegraphics[scale=0.39]{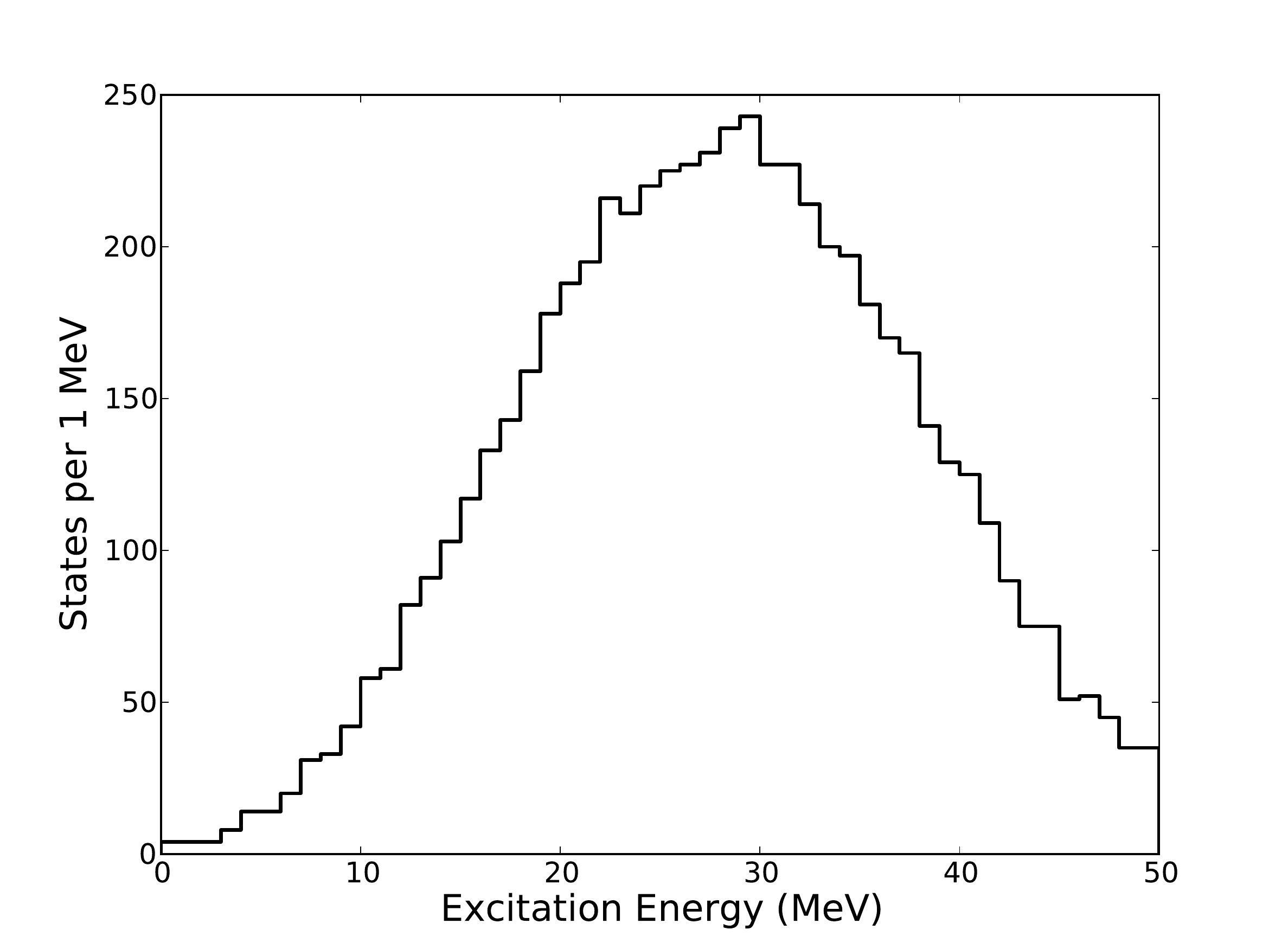}
\caption{$^{28}$Na density of states.  Model space restrictions make the density of states very low, with the departure from exponential growth occurring at low excitation.}
\label{fig:na_dos}
\end{figure}

\begin{figure}
\centering
\includegraphics[scale=0.33]{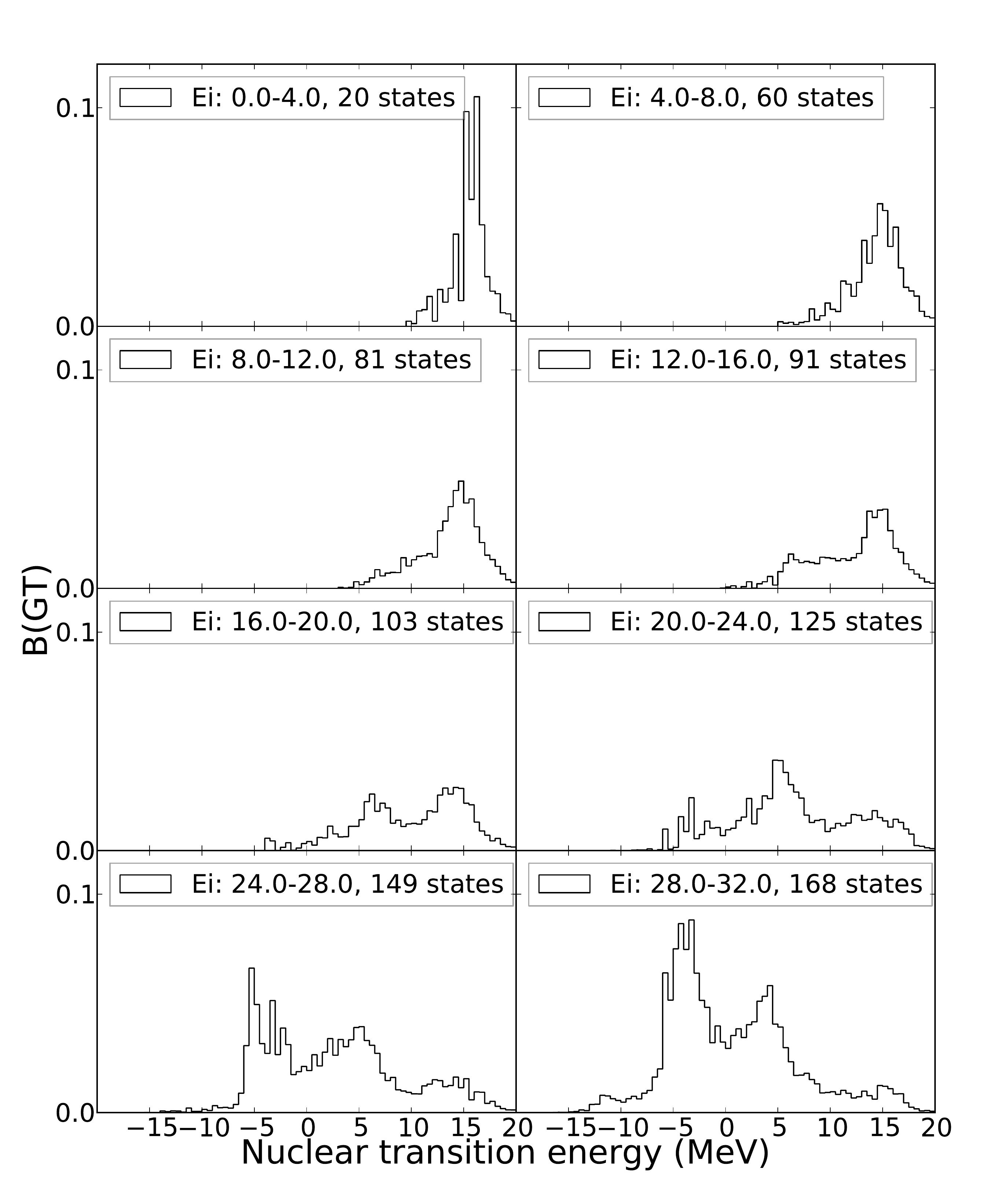}
\caption{Strength distribution for $^{28}$Na.  There is no obvious energy regime where the strength is independent of initial excitation.}
\label{fig:na_str}
\end{figure}

With only 3 protons and 3 neutron holes in the sd shell, $^{28}$Na really pushes the limits of the model space; we see in figure \ref{fig:na_dos} that the density of states is very low, with the inflection point at 12 MeV or less.  While the sd shell may yield acceptable results for low-lying states in $^{28}$Na, we cannot rely on it to get the high initial excitation distributions.  Figure \ref{fig:na_str} shows that there is essentially no initial energy region with a stable strength distribution, as per expectation from the low-lying inflection point in the density of states, though we might speculate at some observable stability between the $E_i=4-8$ MeV and $8-12$ MeV bins.  This observation coupled with the results from the other nuclei make it clear that when model space restrictions severely limit the density of states, the strength distribution is not independent of initial excitation energy.

\section{Computation of Transition Rate}
\label{sec:rate}
Throughout this section, we will use natural units such that $\hbar=c=k_B=1$.  Following FFNI, the electron capture rate for a given initial nuclear state is
\begin{equation}
\lambda_{if}=\mathrm{ln}(2)\frac{f_{if}(T,\mu_e)}{(ft)_{if}}
\end{equation}
where $(ft)_{if}$ is the ft-value appropriate for the transition from parent nucleus state $i$ to daughter nucleus state $f$.  Here, $(ft)_{if}$ is computed from the corresponding Gamow-Teller ($M^{GT}_{if}$) and Fermi ($M^{F}_{if}$) matrix elements by
\begin{eqnarray}
\mathrm{log}(ft^{GT}_{if})=3.596-\mathrm{log}(\vert M^{GT}_{if}\vert^2) \\
\mathrm{log}(ft^{F}_{if})=3.791-\mathrm{log}(\vert M^{F}_{if}\vert^2) \\
\frac{1}{(ft)_{if}}=\frac{1}{ft^{GT}_{if}}+\frac{1}{ft^F_{if}}.
\end{eqnarray}
The factor $f_{if}(T,\mu_e)$ is the phase space integral for the incoming electron and outgoing neutrino.  T is the temperature, and $\mu_e$ is the electron Fermi energy, including rest mass.  The numerical values ``3.596'' and ``3.791'' correspond to choices of axial vector and vector couplings chosen to match those used in FFN, to facilitate comparison.  The phase space integral is
\begin{equation}
f_{if}=\int_{w_l}^\infty w^2(w-q)^2G(Z,w)f_e(w,\mu_e,T)(1-f_\nu)dw
\end{equation}
where w is the total electron energy in units of electron mass, q is the change in total nuclear energy $M_f+E_f-M_i-E_i$ in units of electron mass, Z is the nuclear charge, and f$_e$ and f$_\nu$ are the electron and neutrino occupation probabilities.  The lower limit w$_l$ is a function of q, as the incoming electron must supply enough energy to the nucleus to make the transition; if q$<$1, then w$_l$=1 (corresponding to zero electron kinetic energy), while if q$>$1, w$_l$=q.  G is related to the Coulomb barrier factor and is detailed in FFNI; rather than use the limiting approximations described in that work, we use the form given by eqn. 5b therein.  Note that that work defines q in the negative sense of its use here; that is to say, q in that work is defined as the parent energy minus the daughter energy.

Up until neutrino trapping sets in at $\rho\sim 10^{12}$ g/cm$^3$, we may take f$_\nu\approx 0$.  Here f$_e(w,\mu_e,T)$ is the Fermi-Dirac distribution $(1+e^{(wm_e-\mu_e)/T})^{-1}$.  Using this and our definition of w$_l$ and integrating over final states, we at last arrive at
\begin{widetext}
\begin{eqnarray}
\lambda_i=\mathrm{ln}(2)\int_{-\infty}^1\left(\frac{B^{GT}_i(q)}{10^{3.596}}+\frac{B^F_i(q)}{10^{3.791}}\right)dq\int_1^\infty f_e(w,\mu_e,T)w^2(w-q)^2G(Z,w)dw \nonumber \\
+\mathrm{ln}(2)\int_1^\infty \left(\frac{B^{GT}_i(q)}{10^{3.596}}+\frac{B^F_i(q)}{10^{3.791}}\right)dq\int_q^\infty f_e(w,\mu_e,T)w^2(w-q)^2G(Z,w)dw
\label{eqn:initial_rate}
\end{eqnarray}
\end{widetext}
where $B^{GT}_i(q)\equiv \sum_{f\in \{q\}}\vert M^{GT}_{if}\vert^2$ and $B^{F}_i(q)\equiv \sum_{f\in \{q\}}\vert M^{F}_{if}\vert^2$, where the sums are over final states $f$ with dimensionless (units of electron mass) Q-value $q$.  To compute the total capture rate, we sum over population index weighted initial states.
\begin{equation}
\Lambda=\sum\limits_i\lambda_i\frac{(2J_i+1)e^{-E_i/T}}{G(T)}
\label{eqn:total_rate}
\end{equation}
where G is the partition function.  Recall, however, that above $\sim$12 MeV, the strength distributions look similar.  Therefore, we propose a modification to the GT Brink-Axel hypothesis by applying a cutoff energy below which all states are included and weighted by their population index, and with all remaining statistical weight carried by a single high energy average state.  This is in contrast to the FFN approach of placing the bulk of the strength in a single resonant transition that is identical for all states.  In other words, where FFN treated all states as having an identical giant GT resonance, we treat all states above the cutoff energy as having exactly the same distribution.

\begin{figure}[here]
\centering
\includegraphics[scale=0.6]{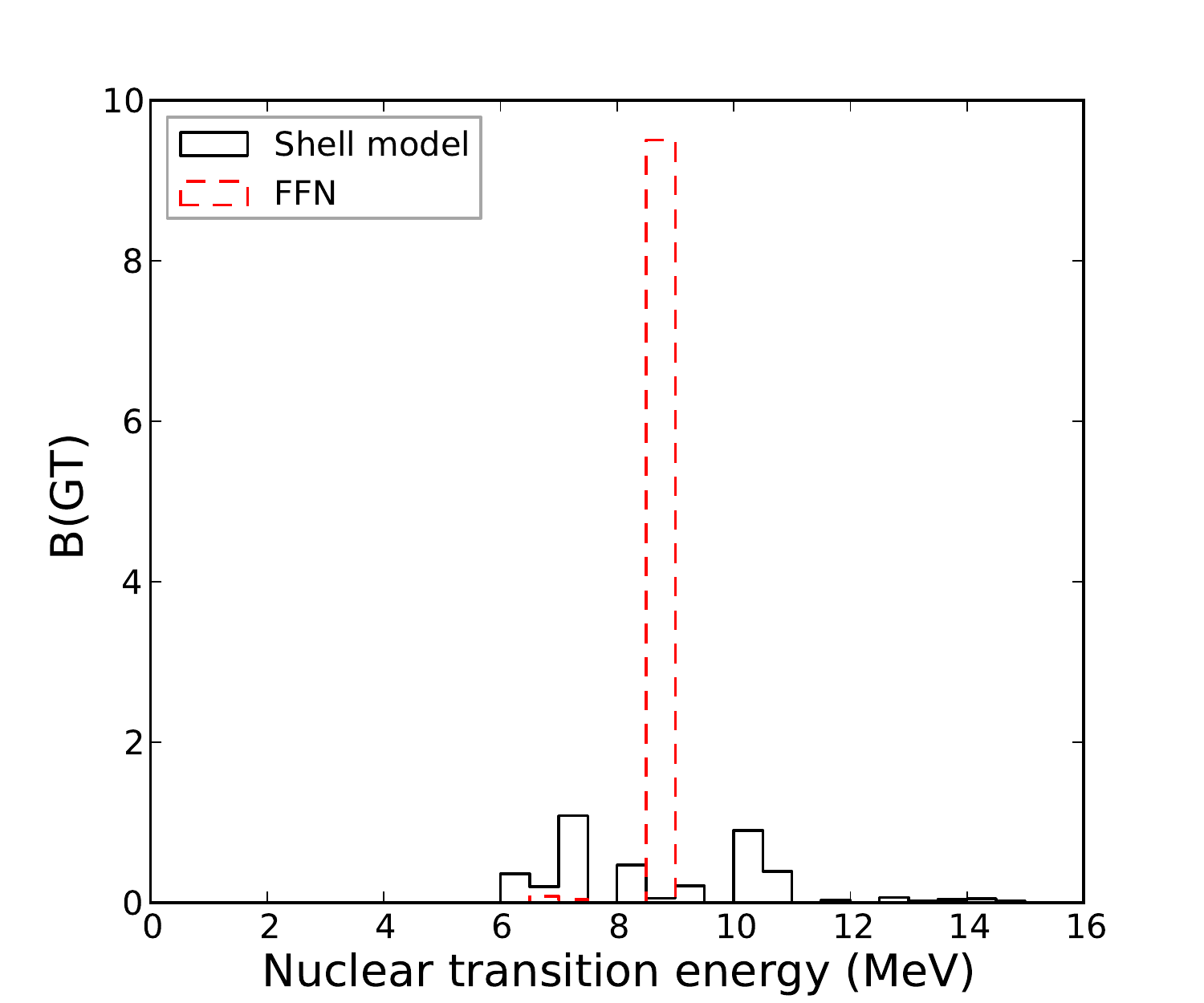}
\caption{$^{28}$Si ground state strength distribution.  The solid line shows the distribution using our shell model calculations, and the dotted line shows the strength from the FFN prescription.  The large peak in the FFN distribution is the GT resonance used in those works.}
\label{fig:si_ground_ffn-ox}
\end{figure}

The difference in these two treatments is profound; figure \ref{fig:si_ground_ffn-ox} shows the strength distributions in the ground state for the FFN approach and the shell model; the large peak in the FFN distribution is the GT resonance.  The two major differences are that the shell model result has less total strength, and the strength is spread to lower transition energies; the former will have the effect of decreasing the capture rate, while the latter will tend to increase it.

Despite the overestimate of the total strength and the misplacement of the resonance, the power of FFN is that it used experimental strengths wherever they were available, and any other technique of computing rates would be well-served by following that example.  Therefore, the strength distributions that we ultimately use to compute capture rates are defined as follows.  (1) We take the same experimentally measured strength distribution as FFN.  (2) We sum the experimental strength and (3) remove that much total strength from our corresponding shell model state by subtracting an equal amount of the lowest-lying strength in the shell model distribution.  (In this procedure we do not correct for quenching.)  (4) We then sum the experimental distribution with what remains of the shell model distribution for that state.  This gives a better estimate of both the capture strength sum rule and the (non-experimental) strength distribution, but is only applicable to initial states with experimentally measured energies.  For transitions from higher, unmeasured initial states, we simply used our shell model distributions; we do not include any shell model parent states with an excitation energy lower than the highest used experimental state.

We now require the nuclear partition function to obtain appropriate initial state occupation indexes.  There are a few approaches to the partition function problem, but in our case, the simplest and most self-consistent is to include only the sd shell states, {\it i.e.} only include in the partition function those states that can be constructed from configurations in the sd shell.  The biggest weaknesses of this approach are that at high enough energies, the density of shell model states actually {\it decreases} to zero, and all negative parity states are neglected, as well as any other states that include configurations with one or more particles promoted into or out of the sd shell.  By the same token, those states will also not be considered to contribute to the electron capture rate, thereby compensating for the overestimate of the included states's occupation indexes.  With the partition function in hand, we can compute the total capture rate from eqn. \ref{eqn:total_rate}.

The electron occupation probability consists of two qualitatively different domains: when $1\leq w\leq\mu_e/m_e$, it varies slowly from a maximum of at most 1 at $w=1$ down to a minimum of 0.5 at $w=\mu_e/m_e$ (we will call this the ``shoulder''), and when $w>\mu_e/m_e$, it is exponentially damped (``tail'').  We numerically integrated the inner integrals of eqn. \ref{eqn:initial_rate} using a combination of two methods, one for each domain.  When the shoulder was part of the integration domain ({\it i.e.}, $q<\mu_e/m_e$), we integrated the shoulder with a 64-point Gauss-Legendre quadrature.  Some or all of the tail is aways in the integration domain, and we integrated it with a 64-point Gauss-Laguerre quadrature.

\begin{figure}[here]
\centering
\includegraphics[scale=0.52]{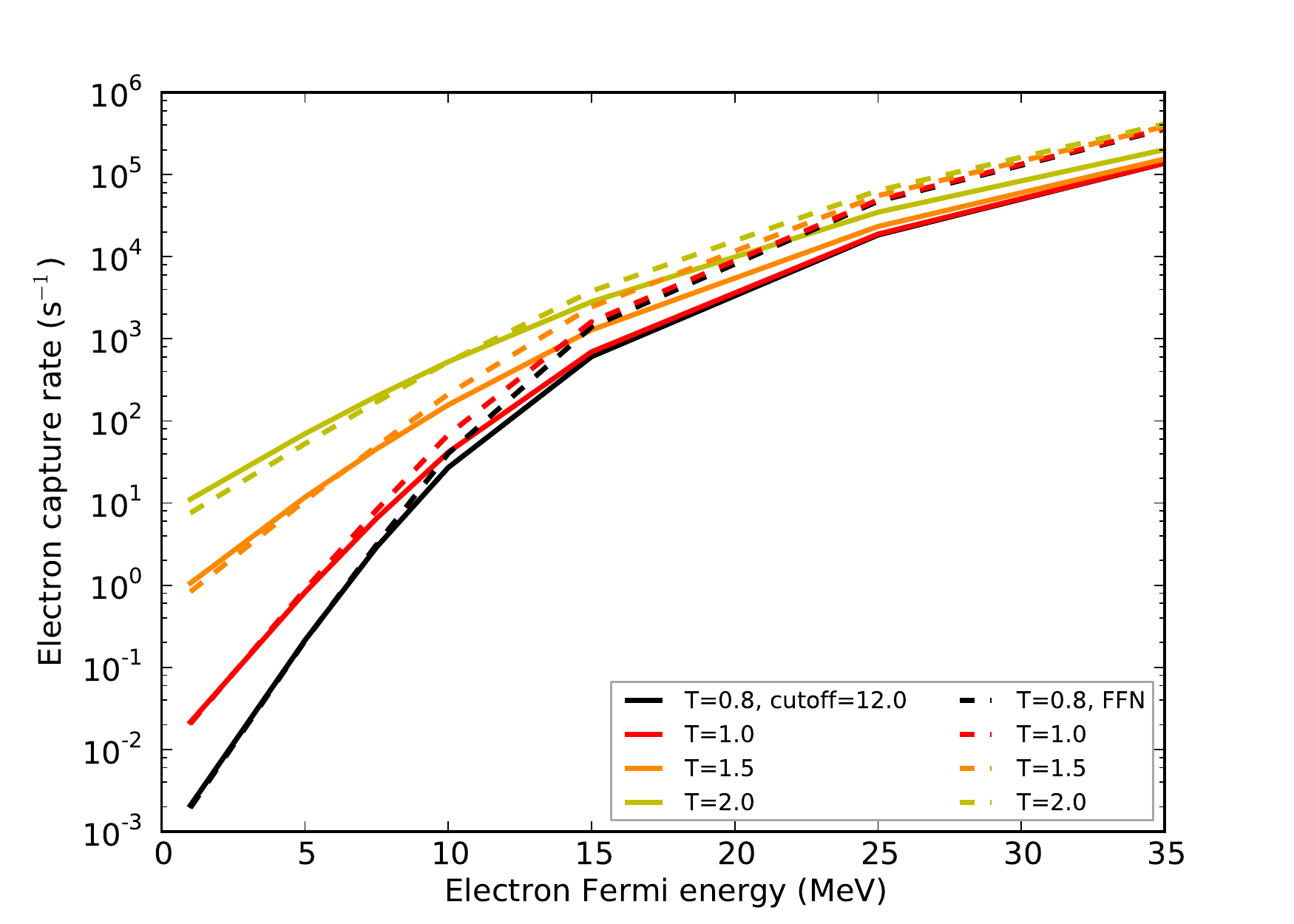}
\caption{Electron capture rates for $^{28}$Si as a function of electron Fermi energy and temperature.  Solid lines show rates with all states up to 12 MeV considered individually and the rest of the statistical weight carried by a single high energy average state, while dashed lines correspond to the rates when all states are assumed to have the same narrow GT resonance, in accordance with the FFN approach to the GT Brink-Axel hypothesis.}
\label{fig:si_rate_hybrid-ffn}
\end{figure}

Figure \ref{fig:si_rate_hybrid-ffn} shows electron capture rates for $^{28}$Si as a function of electron Fermi energy and temperature.  The solid lines were computed using a cutoff energy of 12 MeV and a high energy average state strength distribution computed from the spin-weighted ($2J+1$) average of every state between 12 and 14 MeV, and the dashed lines are the rates computed using the FFN resonance prescription.  At sufficiently high Fermi energy, there are enough electrons above the GT resonance used in the FFN approach for the rates to outstrip those of our shell model results, as the large amount of strength in the resonance outcompetes the shell model.  However, at low Fermi energy, the spread of strength to low transition energies found in the shell model approach serves to boost the rates above the FFN estimates.

\begin{figure}[here]
\centering
\includegraphics[scale=0.52]{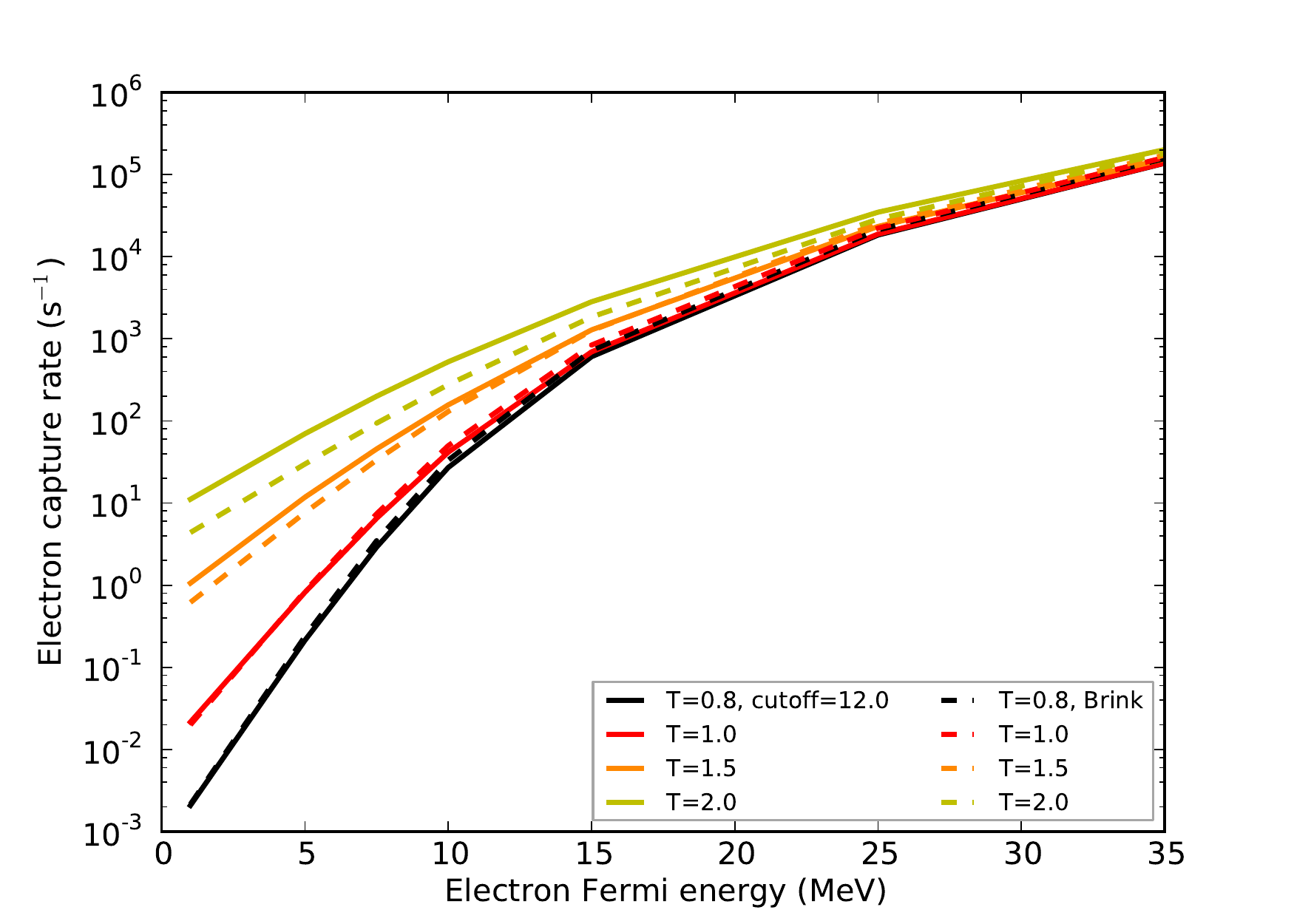}
\caption{Electron capture rates for $^{28}$Si as a function of electron Fermi energy and temperature.  Solid lines show rates with all states up to 12 MeV considered individually and the rest of the statistical weight carried by a single high energy average state, while dashed lines correspond to the rates when all states are assumed to have the same bulk GT strength distribution as our shell model calculation of the ground state.}
\label{fig:si_rate_hybrid-brink}
\end{figure}

Figure \ref{fig:si_rate_hybrid-brink} compares the shell model capture rates with a cutoff of 12 MeV against a GT Brink-Axel approach (as in fig.  \ref{fig:si_rate_hybrid-ffn}), but with the single resonance in the FFN model replaced by the shell model strength distribution for the ground state.  That is, in the ``Brink'' approach here, we used experimental values of the transition strength for each initial state where known, and the rest of the strength in each excited state is carried by the ground state distribution.  In contrast to the behavior of the FFN approach, the shell model Brink-Axel curves lack the marked jump above the more comprehensive shell model rates as Fermi energy increases, and they eventually converge.  It is notable that the GT Brink-Axel results are not uniformly greater {\it or} lesser than the more comprehensive shell model rates; in the Fermi energy region between 5 and 15 MeV, the T=0.8 and 1.0 MeV Brink-Axel rates just peek above the corresponding shell model rates.

\begin{figure}[here]
\centering
\includegraphics[scale=0.52]{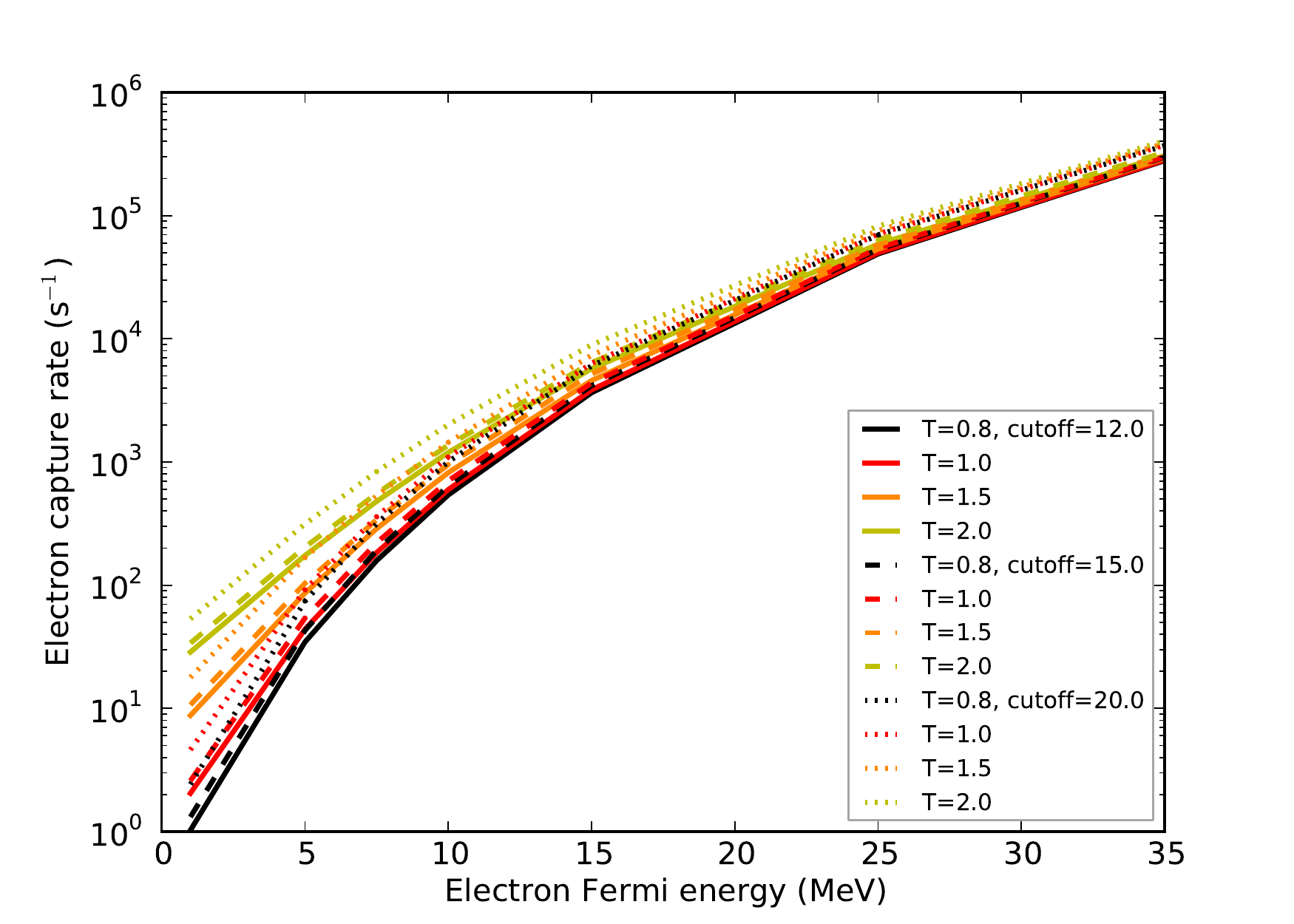}
\caption{Thermodynamically unweighted electron capture rates for high energy average states in $^{28}$Si. The solid lines are the rates for an HEA state with a cutoff energy of 12 MeV, the dashed lines show a cutoff of 15 MeV, and the dotted lines are for a cutoff of 20 MeV.}
\label{fig:si_rate_hea}
\end{figure}

In light of the apparent sensitivity to how excited states are handled in rate calculations, we compare in figure \ref{fig:si_rate_hea} the \emph{thermodynamically unweighted} (meaning the population factor is not included) capture rates of the high energy average states corresponding to several cutoff energies (the HEA state being that which carries all of the statistical weight above the cuttoff).  The solid lines show rates for an HEA state including all shell model states between 12 and 14 MeV, as in the previous calculations.  The dashed lines give the rates of an HEA state computed from all states between 15 and 16 MeV, and the dotted lines are for an HEA state comprised of states between 20 and 20.3 MeV.  The widths for the averaging were chosen such that each HEA state was comprised of at least 50 individual states.

\begin{figure}[here]
\centering
\includegraphics[scale=0.5]{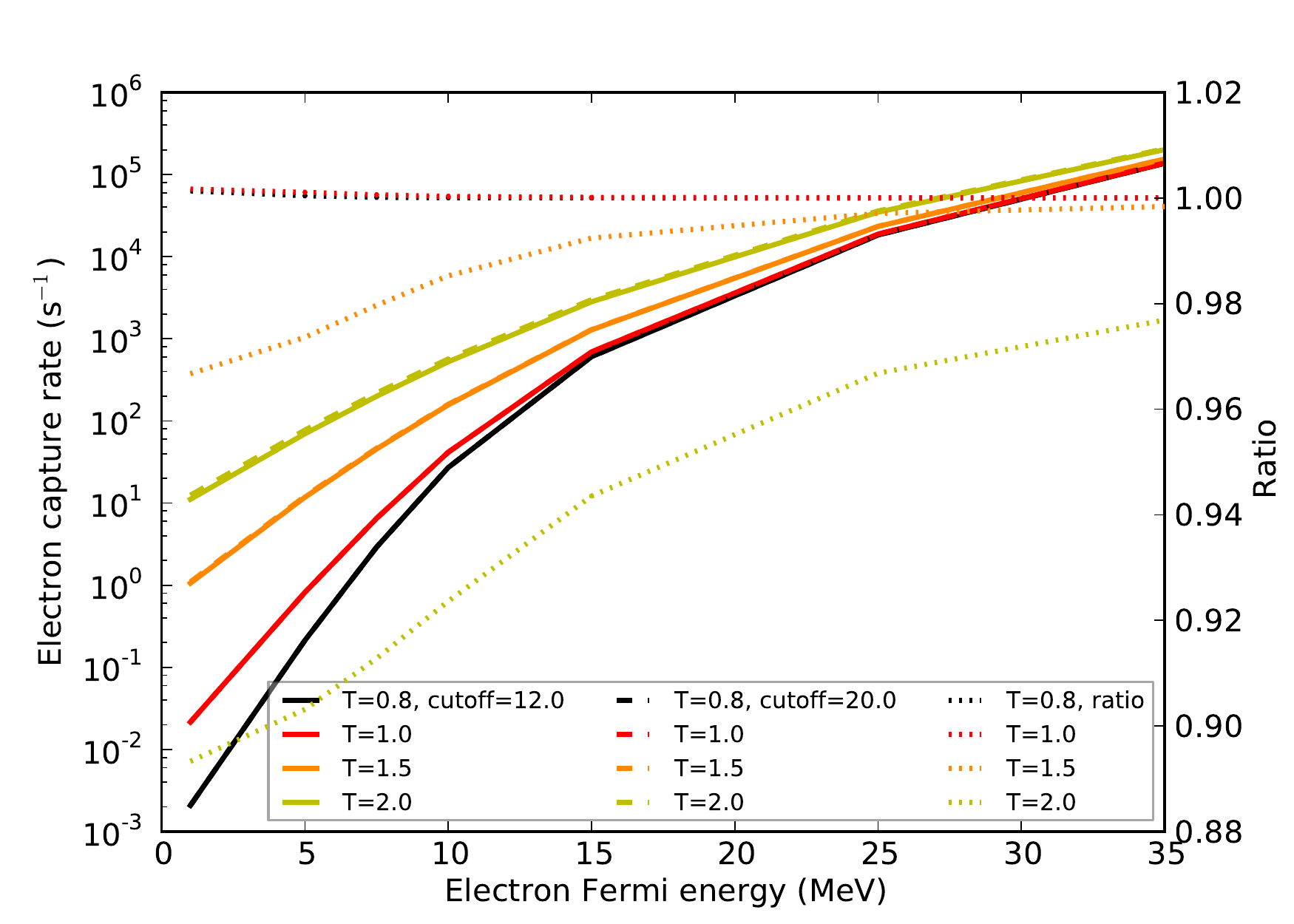}
\caption{Electron capture rates for $^{28}$Si comparing two choices of cutoff energy.  The solid lines correspond to a cutoff energy of 12 MeV, while the dashed lines are for a cutoff of 20 MeV.  The dotted lines show the ratio of the cutoff = 12 rates to the cutoff = 20 rates.  That the rates are nearly identical lends credence to the technique of using a high energy average state.}
\label{fig:si_rate_hybrid-hybrid}
\end{figure}

The rates for all three HEA states differ from one another by at the most a factor of 3 in the range considered, which is offset by the reduction in statistical weight carried by the HEA state as the cutoff energy increases.  The HEA statistical weight is simply the remaining probability after the occupation indexes of all lower-energy states are accounted for:
\begin{equation}
w_{HEA}=1-\frac{1}{G(T)}\sum\limits_{E_i<E_{cutoff}}e^{-E_i/T}
\end{equation}
The weights for the given cutoff energies and temperatures are shown in table \ref{table:hea_weights}.  Over most of the temperature range, the weight falls off much faster than the unweighted HEA rate grows with cutoff energy.  Figure \ref{fig:si_rate_hybrid-hybrid} shows the total capture rates for cutoff energies of 12 and 20 MeV as well as the ratio of the low cutoff rates to the high cutoff rates.  Over a broad range of temperature and electron Fermi energy, the two choices produce nearly identical results; up to temperatures of 1.5 MeV, the rates agree to within 3 percent, and differ by only $\sim10$ percent in the extreme case of T$=2$ MeV and E$_f<5$ MeV.  The evidently small errors introduced by a particular choice of cutoff energy will ultimately be washed out by other uncertainties, including the eventual treatment of quenching

\begin{table}[here]
\begin{tabular}{ | c | c | c | c | }
\hline
T (MeV) & Cutoff = 12 MeV & 15 MeV & 20 MeV \\
\hline
0.8 & $2.86\times 10^{-5}$ & $1.70\times 10^{-6}$ & $1.23\times 10^{-8}$ \\
1.0 & $6.30\times 10^{-4}$ & $7.76\times 10^{-5}$ & $1.88\times 10^{-6}$ \\
1.5 & $3.49\times 10^{-2}$ & $1.11\times 10^{-2}$ & $1.31\times 10^{-3}$ \\
2.0 & $2.00\times 10^{-1}$ & $9.94\times 10^{-2}$ & $2.53\times 10^{-2}$ \\
\hline
\end{tabular}
\caption{Statistical weights of the high energy average state as a function of temperature and cutoff energy.}
\label{table:hea_weights}
\end{table}

\section{Discussion and Conclusions}
\label{sec:discussion}
The three principle observations from this work are that 1) at high excitation energies the GT strength distribution does not depend sensitively on nuclear excitation energy (though it is a function of isospin), 2) the GT strength distribution spreads to low and negative transition energies, and 3) the spreading of the strength tends to increase the electron capture rate, as not only does it decrease the electron capture energy threshold, but for a given incoming electron, it also increases the phase space of the outgoing neutrino.

As seen in figures \ref{fig:si_rate_hybrid-ffn} and \ref{fig:si_rate_hybrid-brink}, point 3 above is contradicted in some regimes of temperature and Fermi energy.  In order to understand why the shell model rates sometimes fall short of other approaches, we return to the total strength, {\it i.e.} sum rule, as a function of excitation energy.  Considering first the FFN approach, comparing figures \ref{fig:si_ground_ffn-ox} and \ref{fig:si_tot_str} reveals that the strength in the GT resonance employed by FFN is about twice the average total strength at all excitation energies computed from the shell model, resulting in an overestimate of the capture rate at high Fermi energies.

The sources of the deviations in the shell model Brink-Axel approach are a little more subtle.  There are regimes of temperature and density where the rates derived from our shell model treatment are greater than those derived using the Brink-Axel assumption.  This stems in part from transitions from the parent nucleus to relatively low-lying discrete states in the daughter nucleus.  These low-lying daughter states have more favorable Q-values (see figure \ref{fig:si_str}).  Furthermore, figure \ref{fig:si_tot_str} shows that on average, the total strength increases slowly with parent nucleus excitation energy (roughly, from B(GT)$\sim$4 in the ground state to B(GT)$\sim$5.5 at 30 MeV), further enhancing the rate relative to that of the parent ground state.  But figure \ref{fig:si_tot_str} also shows that at relatively low excitation energies, there are two significant drops in the total strength, which account for the regions where the Brink-Axel rate exceeds the shell model rate.  Recall that the GT Brink-Axel approach treats {\it all} excited states as having the same bulk GT strength distribution as the ground state, but the more comprehensive model includes contributions from those states that have less total strength.  Importantly, some of those states are at low initial excitation.  Hence, they do not have the low-lying strength (low Q-value) seen in higher states, and they have a comparatively large population factor.  The combination of low total strength, no low-lying strength, and a large population factor yield temperature and Fermi energy regimes where the Brink-Axel approach overestimates the rate.

Ultimately, we must conclude that the GT Brink-Axel hypothesis as it has been traditionally used is likely inappropriate for obtaining accurate electron capture rates -- and by extension, all nuclear weak rates -- at the high temperatures and densities characteristic of collapsing supernova cores.  We must be circumspect, however, as the nuclei examined here are very light by supernova core standards.  {\it If} later work is able to demonstrate that the trends found here are applicable to larger nuclei, then we will have found a useful technique for simplifying the accurate computation of weak rates in those nuclei.

The analysis of $^{28}$Si in this work is essentially a cruder version of the work of Oda et al.  We performed no careful matching of the energies of the daughter states relative to the parent states, meaning that where experimental data were not used, the distributions shown here will not have precise transition energies.  This imprecision is unimportant for the sake of our goal here, which was to demonstrate the failure of the GT Brink-Axel hypothesis and how it can be modified for use at high initial excitation.  With these results and the 20 or so years of experimental data collected since the Oda et al rate survey, though, it is worth re-examining the weak rate calculations for sd-shell nuclei, which are important in the late phases of stellar evolution leading up to core collapse.

This leaves us with two major directions to follow up.   First, we will recompute the weak rates for all sd-shell nuclei over a wide range of temperatures and densities relevant to late stellar evolution and core collapse using our modification to the GT Brink-Axel hypothesis and the most recent experimental data.  Second, we will seek ways to extend the results presented in this paper to the large, neutron-rich nuclei that are abundant during collapse.

\section{Acknowledgements}
This work was supported in part by NSF Grant No. PHY-1307372 at UCSD and NSF Grant No. PHY-1404442 at MSU.  We would like to thank Calvin Johnson for helpful discussions.

\bibliography{../references.bib}

\begin{thebibliography}{58}
\expandafter\ifx\csname natexlab\endcsname\relax\def\natexlab#1{#1}\fi
\expandafter\ifx\csname bibnamefont\endcsname\relax
  \def\bibnamefont#1{#1}\fi
\expandafter\ifx\csname bibfnamefont\endcsname\relax
  \def\bibfnamefont#1{#1}\fi
\expandafter\ifx\csname citenamefont\endcsname\relax
  \def\citenamefont#1{#1}\fi
\expandafter\ifx\csname url\endcsname\relax
  \def\url#1{\texttt{#1}}\fi
\expandafter\ifx\csname urlprefix\endcsname\relax\def\urlprefix{URL }\fi
\providecommand{\bibinfo}[2]{#2}
\providecommand{\eprint}[2][]{\url{#2}}

\bibitem[{\citenamefont{Brink}(1955)}]{brink:thesis}
\bibinfo{author}{\bibfnamefont{D.~M.} \bibnamefont{Brink}}, Ph.D. thesis,
  \bibinfo{school}{Oxford University} (\bibinfo{year}{1955}).

\bibitem[{\citenamefont{{Axel}}(1962)}]{axel:1962}
\bibinfo{author}{\bibfnamefont{P.}~\bibnamefont{{Axel}}},
  \bibinfo{journal}{Physical Review} \textbf{\bibinfo{volume}{126}},
  \bibinfo{pages}{671} (\bibinfo{year}{1962}).

\bibitem[{\citenamefont{{Szefli{\'n}ski}
  et~al.}(1983)\citenamefont{{Szefli{\'n}ski}, {Szefli{\'n}ska}, {Wilhelmi},
  {Rz\c aca-Urban}, {Klapdor}, {Anderson}, {Grotz}, and
  {Metzinger}}}]{szeflinski-etal:1983}
\bibinfo{author}{\bibfnamefont{Z.}~\bibnamefont{{Szefli{\'n}ski}}},
  \bibinfo{author}{\bibfnamefont{G.}~\bibnamefont{{Szefli{\'n}ska}}},
  \bibinfo{author}{\bibfnamefont{Z.}~\bibnamefont{{Wilhelmi}}},
  \bibinfo{author}{\bibfnamefont{T.}~\bibnamefont{{Rz\c aca-Urban}}},
  \bibinfo{author}{\bibfnamefont{H.~V.} \bibnamefont{{Klapdor}}},
  \bibinfo{author}{\bibfnamefont{E.}~\bibnamefont{{Anderson}}},
  \bibinfo{author}{\bibfnamefont{K.}~\bibnamefont{{Grotz}}}, \bibnamefont{and}
  \bibinfo{author}{\bibfnamefont{J.}~\bibnamefont{{Metzinger}}},
  \bibinfo{journal}{Physics Letters B} \textbf{\bibinfo{volume}{126}},
  \bibinfo{pages}{159} (\bibinfo{year}{1983}).

\bibitem[{\citenamefont{{Fuller} et~al.}(1980)\citenamefont{{Fuller}, {Fowler},
  and {Newman}}}]{ffn:1980}
\bibinfo{author}{\bibfnamefont{G.~M.} \bibnamefont{{Fuller}}},
  \bibinfo{author}{\bibfnamefont{W.~A.} \bibnamefont{{Fowler}}},
  \bibnamefont{and} \bibinfo{author}{\bibfnamefont{M.~J.}
  \bibnamefont{{Newman}}}, \bibinfo{journal}{Astrophys. J. (Supplement)}
  \textbf{\bibinfo{volume}{42}}, \bibinfo{pages}{447} (\bibinfo{year}{1980}).

\bibitem[{\citenamefont{{Aufderheide} et~al.}(1994)\citenamefont{{Aufderheide},
  {Fushiki}, {Woosley}, and {Hartmann}}}]{afwh:1994}
\bibinfo{author}{\bibfnamefont{M.~B.} \bibnamefont{{Aufderheide}}},
  \bibinfo{author}{\bibfnamefont{I.}~\bibnamefont{{Fushiki}}},
  \bibinfo{author}{\bibfnamefont{S.~E.} \bibnamefont{{Woosley}}},
  \bibnamefont{and} \bibinfo{author}{\bibfnamefont{D.~H.}
  \bibnamefont{{Hartmann}}}, \bibinfo{journal}{Astrophys. J. (Supplement)}
  \textbf{\bibinfo{volume}{91}}, \bibinfo{pages}{389} (\bibinfo{year}{1994}).

\bibitem[{\citenamefont{{Kar} et~al.}(1994)\citenamefont{{Kar}, {Ray}, and
  {Sarkar}}}]{krs:1994}
\bibinfo{author}{\bibfnamefont{K.}~\bibnamefont{{Kar}}},
  \bibinfo{author}{\bibfnamefont{A.}~\bibnamefont{{Ray}}}, \bibnamefont{and}
  \bibinfo{author}{\bibfnamefont{S.}~\bibnamefont{{Sarkar}}},
  \bibinfo{journal}{\apj} \textbf{\bibinfo{volume}{434}}, \bibinfo{pages}{662}
  (\bibinfo{year}{1994}).

\bibitem[{\citenamefont{{Chakravarti} et~al.}(1999)\citenamefont{{Chakravarti},
  {Kar}, {Ray}, and {Sarkar}}}]{ckrs:1999}
\bibinfo{author}{\bibfnamefont{S.}~\bibnamefont{{Chakravarti}}},
  \bibinfo{author}{\bibfnamefont{K.}~\bibnamefont{{Kar}}},
  \bibinfo{author}{\bibfnamefont{A.}~\bibnamefont{{Ray}}}, \bibnamefont{and}
  \bibinfo{author}{\bibfnamefont{S.}~\bibnamefont{{Sarkar}}},
  \bibinfo{journal}{ArXiv Astrophysics e-prints}  (\bibinfo{year}{1999}),
  \eprint{astro-ph/9910058}.

\bibitem[{\citenamefont{{Kajino} et~al.}(1988)\citenamefont{{Kajino}, {Shiino},
  {Toki}, {Brown}, and {Wildenthal}}}]{kajino-etal:1988}
\bibinfo{author}{\bibfnamefont{T.}~\bibnamefont{{Kajino}}},
  \bibinfo{author}{\bibfnamefont{E.}~\bibnamefont{{Shiino}}},
  \bibinfo{author}{\bibfnamefont{H.}~\bibnamefont{{Toki}}},
  \bibinfo{author}{\bibfnamefont{B.~A.} \bibnamefont{{Brown}}},
  \bibnamefont{and} \bibinfo{author}{\bibfnamefont{B.~H.}
  \bibnamefont{{Wildenthal}}}, \bibinfo{journal}{Nuclear Physics A}
  \textbf{\bibinfo{volume}{480}}, \bibinfo{pages}{175} (\bibinfo{year}{1988}).

\bibitem[{\citenamefont{{Langanke} and {Martinez-Pinedo}}(1998)}]{lm:1998}
\bibinfo{author}{\bibfnamefont{K.}~\bibnamefont{{Langanke}}} \bibnamefont{and}
  \bibinfo{author}{\bibfnamefont{G.}~\bibnamefont{{Martinez-Pinedo}}},
  \bibinfo{journal}{Physics Letters B} \textbf{\bibinfo{volume}{436}},
  \bibinfo{pages}{19} (\bibinfo{year}{1998}), \eprint{nucl-th/9809081}.

\bibitem[{\citenamefont{{Caurier} et~al.}(1999)\citenamefont{{Caurier},
  {Langanke}, {Mart{\'{\i}}nez-Pinedo}, and {Nowacki}}}]{clmn:1999}
\bibinfo{author}{\bibfnamefont{E.}~\bibnamefont{{Caurier}}},
  \bibinfo{author}{\bibfnamefont{K.}~\bibnamefont{{Langanke}}},
  \bibinfo{author}{\bibfnamefont{G.}~\bibnamefont{{Mart{\'{\i}}nez-Pinedo}}},
  \bibnamefont{and}
  \bibinfo{author}{\bibfnamefont{F.}~\bibnamefont{{Nowacki}}},
  \bibinfo{journal}{Nuclear Physics A} \textbf{\bibinfo{volume}{653}},
  \bibinfo{pages}{439} (\bibinfo{year}{1999}), \eprint{nucl-th/9903042}.

\bibitem[{\citenamefont{{Langanke} and
  {Mart{\'{\i}}nez-Pinedo}}(1999)}]{lm:1999}
\bibinfo{author}{\bibfnamefont{K.}~\bibnamefont{{Langanke}}} \bibnamefont{and}
  \bibinfo{author}{\bibfnamefont{G.}~\bibnamefont{{Mart{\'{\i}}nez-Pinedo}}},
  \bibinfo{journal}{Physics Letters B} \textbf{\bibinfo{volume}{453}},
  \bibinfo{pages}{187} (\bibinfo{year}{1999}), \eprint{nucl-th/9809082}.

\bibitem[{\citenamefont{{Martinezpinedo} and {Langanke}}(1999)}]{ml:1999}
\bibinfo{author}{\bibfnamefont{G.}~\bibnamefont{{Martinezpinedo}}}
  \bibnamefont{and}
  \bibinfo{author}{\bibfnamefont{K.}~\bibnamefont{{Langanke}}},
  \bibinfo{journal}{Nuclear Physics A} \textbf{\bibinfo{volume}{654}},
  \bibinfo{pages}{904} (\bibinfo{year}{1999}).

\bibitem[{\citenamefont{{Langanke} and
  {Mart{\'{\i}}nez-Pinedo}}(2000)}]{lm:2000}
\bibinfo{author}{\bibfnamefont{K.}~\bibnamefont{{Langanke}}} \bibnamefont{and}
  \bibinfo{author}{\bibfnamefont{G.}~\bibnamefont{{Mart{\'{\i}}nez-Pinedo}}},
  \bibinfo{journal}{Nuclear Physics A} \textbf{\bibinfo{volume}{673}},
  \bibinfo{pages}{481} (\bibinfo{year}{2000}), \eprint{nucl-th/0001018}.

\bibitem[{\citenamefont{{Langanke} and
  {Mart{\'{\i}}nez-Pinedo}}(2001)}]{lm:2001}
\bibinfo{author}{\bibfnamefont{K.}~\bibnamefont{{Langanke}}} \bibnamefont{and}
  \bibinfo{author}{\bibfnamefont{G.}~\bibnamefont{{Mart{\'{\i}}nez-Pinedo}}},
  \bibinfo{journal}{Atomic Data and Nuclear Data Tables}
  \textbf{\bibinfo{volume}{79}}, \bibinfo{pages}{1} (\bibinfo{year}{2001}).

\bibitem[{\citenamefont{{Cole} et~al.}(2012)\citenamefont{{Cole}, {Anderson},
  {Zegers}, {Austin}, {Brown}, {Valdez}, {Gupta}, {Hitt}, and
  {Fawwaz}}}]{cole-etal:2012}
\bibinfo{author}{\bibfnamefont{A.~L.} \bibnamefont{{Cole}}},
  \bibinfo{author}{\bibfnamefont{T.~S.} \bibnamefont{{Anderson}}},
  \bibinfo{author}{\bibfnamefont{R.~G.~T.} \bibnamefont{{Zegers}}},
  \bibinfo{author}{\bibfnamefont{S.~M.} \bibnamefont{{Austin}}},
  \bibinfo{author}{\bibfnamefont{B.~A.} \bibnamefont{{Brown}}},
  \bibinfo{author}{\bibfnamefont{L.}~\bibnamefont{{Valdez}}},
  \bibinfo{author}{\bibfnamefont{S.}~\bibnamefont{{Gupta}}},
  \bibinfo{author}{\bibfnamefont{G.~W.} \bibnamefont{{Hitt}}},
  \bibnamefont{and} \bibinfo{author}{\bibfnamefont{O.}~\bibnamefont{{Fawwaz}}},
  \bibinfo{journal}{\prc} \textbf{\bibinfo{volume}{86}}, \bibinfo{eid}{015809}
  (\bibinfo{year}{2012}), \eprint{1204.1994}.

\bibitem[{\citenamefont{{Fischer} et~al.}(2013)\citenamefont{{Fischer},
  {Langanke}, and {Mart{\'{\i}}nez-Pinedo}}}]{flm:2013}
\bibinfo{author}{\bibfnamefont{T.}~\bibnamefont{{Fischer}}},
  \bibinfo{author}{\bibfnamefont{K.}~\bibnamefont{{Langanke}}},
  \bibnamefont{and}
  \bibinfo{author}{\bibfnamefont{G.}~\bibnamefont{{Mart{\'{\i}}nez-Pinedo}}},
  \bibinfo{journal}{\prc} \textbf{\bibinfo{volume}{88}}, \bibinfo{eid}{065804}
  (\bibinfo{year}{2013}), \eprint{1309.4271}.

\bibitem[{\citenamefont{{Langanke} et~al.}(2001)\citenamefont{{Langanke},
  {Kolbe}, and {Dean}}}]{lkd:2001}
\bibinfo{author}{\bibfnamefont{K.}~\bibnamefont{{Langanke}}},
  \bibinfo{author}{\bibfnamefont{E.}~\bibnamefont{{Kolbe}}}, \bibnamefont{and}
  \bibinfo{author}{\bibfnamefont{D.~J.} \bibnamefont{{Dean}}},
  \bibinfo{journal}{\prc} \textbf{\bibinfo{volume}{63}}, \bibinfo{eid}{032801}
  (\bibinfo{year}{2001}), \eprint{nucl-th/0012036}.

\bibitem[{\citenamefont{{Mart{\'{\i}}nez-Pinedo}
  et~al.}(2014)\citenamefont{{Mart{\'{\i}}nez-Pinedo}, {Lam}, {Langanke},
  {Zegers}, and {Sullivan}}}]{martinez-pinedo-etal:2014}
\bibinfo{author}{\bibfnamefont{G.}~\bibnamefont{{Mart{\'{\i}}nez-Pinedo}}},
  \bibinfo{author}{\bibfnamefont{Y.~H.} \bibnamefont{{Lam}}},
  \bibinfo{author}{\bibfnamefont{K.}~\bibnamefont{{Langanke}}},
  \bibinfo{author}{\bibfnamefont{R.~G.~T.} \bibnamefont{{Zegers}}},
  \bibnamefont{and}
  \bibinfo{author}{\bibfnamefont{C.}~\bibnamefont{{Sullivan}}},
  \bibinfo{journal}{\prc} \textbf{\bibinfo{volume}{89}}, \bibinfo{eid}{045806}
  (\bibinfo{year}{2014}), \eprint{1402.0793}.

\bibitem[{\citenamefont{{Juodagalvis} et~al.}(2010)\citenamefont{{Juodagalvis},
  {Langanke}, {Hix}, {Mart{\'{\i}}nez-Pinedo}, and
  {Sampaio}}}]{juodagalvis-etal:2010}
\bibinfo{author}{\bibfnamefont{A.}~\bibnamefont{{Juodagalvis}}},
  \bibinfo{author}{\bibfnamefont{K.}~\bibnamefont{{Langanke}}},
  \bibinfo{author}{\bibfnamefont{W.~R.} \bibnamefont{{Hix}}},
  \bibinfo{author}{\bibfnamefont{G.}~\bibnamefont{{Mart{\'{\i}}nez-Pinedo}}},
  \bibnamefont{and} \bibinfo{author}{\bibfnamefont{J.~M.}
  \bibnamefont{{Sampaio}}}, \bibinfo{journal}{Nuclear Physics A}
  \textbf{\bibinfo{volume}{848}}, \bibinfo{pages}{454} (\bibinfo{year}{2010}),
  \eprint{0909.0179}.

\bibitem[{\citenamefont{{Arnett}}(1977)}]{arnett:1977}
\bibinfo{author}{\bibfnamefont{W.~D.} \bibnamefont{{Arnett}}},
  \bibinfo{journal}{\apj} \textbf{\bibinfo{volume}{218}}, \bibinfo{pages}{815}
  (\bibinfo{year}{1977}).

\bibitem[{\citenamefont{{Bowers} and {Wilson}}(1982)}]{bw:1982}
\bibinfo{author}{\bibfnamefont{R.}~\bibnamefont{{Bowers}}} \bibnamefont{and}
  \bibinfo{author}{\bibfnamefont{J.~R.} \bibnamefont{{Wilson}}},
  \bibinfo{journal}{\apj} \textbf{\bibinfo{volume}{263}}, \bibinfo{pages}{366}
  (\bibinfo{year}{1982}).

\bibitem[{\citenamefont{{Bethe} and {Wilson}}(1985)}]{bw:1985}
\bibinfo{author}{\bibfnamefont{H.~A.} \bibnamefont{{Bethe}}} \bibnamefont{and}
  \bibinfo{author}{\bibfnamefont{J.~R.} \bibnamefont{{Wilson}}},
  \bibinfo{journal}{\apj} \textbf{\bibinfo{volume}{295}}, \bibinfo{pages}{14}
  (\bibinfo{year}{1985}).

\bibitem[{\citenamefont{{Blondin} et~al.}(2003)\citenamefont{{Blondin},
  {Mezzacappa}, and {DeMarino}}}]{bmd:2003}
\bibinfo{author}{\bibfnamefont{J.~M.} \bibnamefont{{Blondin}}},
  \bibinfo{author}{\bibfnamefont{A.}~\bibnamefont{{Mezzacappa}}},
  \bibnamefont{and}
  \bibinfo{author}{\bibfnamefont{C.}~\bibnamefont{{DeMarino}}},
  \bibinfo{journal}{\apj} \textbf{\bibinfo{volume}{584}}, \bibinfo{pages}{971}
  (\bibinfo{year}{2003}).

\bibitem[{\citenamefont{{Blondin} and {Mezzacappa}}(2007)}]{bm:2007}
\bibinfo{author}{\bibfnamefont{J.~M.} \bibnamefont{{Blondin}}}
  \bibnamefont{and}
  \bibinfo{author}{\bibfnamefont{A.}~\bibnamefont{{Mezzacappa}}},
  \bibinfo{journal}{\nat} \textbf{\bibinfo{volume}{445}}, \bibinfo{pages}{58}
  (\bibinfo{year}{2007}), \eprint{arXiv:astro-ph/0611680}.

\bibitem[{\citenamefont{{Scheck} et~al.}(2008)\citenamefont{{Scheck}, {Janka},
  {Foglizzo}, and {Kifonidis}}}]{sjfk:2008}
\bibinfo{author}{\bibfnamefont{L.}~\bibnamefont{{Scheck}}},
  \bibinfo{author}{\bibfnamefont{H.-T.} \bibnamefont{{Janka}}},
  \bibinfo{author}{\bibfnamefont{T.}~\bibnamefont{{Foglizzo}}},
  \bibnamefont{and}
  \bibinfo{author}{\bibfnamefont{K.}~\bibnamefont{{Kifonidis}}},
  \bibinfo{journal}{Astronomy and Astrophysics} \textbf{\bibinfo{volume}{477}},
  \bibinfo{pages}{931} (\bibinfo{year}{2008}), \eprint{0704.3001}.

\bibitem[{\citenamefont{{Brandt} et~al.}(2011)\citenamefont{{Brandt},
  {Burrows}, {Ott}, and {Livne}}}]{bbol:2011}
\bibinfo{author}{\bibfnamefont{T.~D.} \bibnamefont{{Brandt}}},
  \bibinfo{author}{\bibfnamefont{A.}~\bibnamefont{{Burrows}}},
  \bibinfo{author}{\bibfnamefont{C.~D.} \bibnamefont{{Ott}}}, \bibnamefont{and}
  \bibinfo{author}{\bibfnamefont{E.}~\bibnamefont{{Livne}}},
  \bibinfo{journal}{\apj} \textbf{\bibinfo{volume}{728}}, \bibinfo{eid}{8}
  (\bibinfo{year}{2011}), \eprint{1009.4654}.

\bibitem[{\citenamefont{{Arcones} et~al.}(2007)\citenamefont{{Arcones},
  {Janka}, and {Scheck}}}]{ajs:2007}
\bibinfo{author}{\bibfnamefont{A.}~\bibnamefont{{Arcones}}},
  \bibinfo{author}{\bibfnamefont{H.-T.} \bibnamefont{{Janka}}},
  \bibnamefont{and} \bibinfo{author}{\bibfnamefont{L.}~\bibnamefont{{Scheck}}},
  \bibinfo{journal}{Astronomy and Astrophysics} \textbf{\bibinfo{volume}{467}},
  \bibinfo{pages}{1227} (\bibinfo{year}{2007}).

\bibitem[{\citenamefont{{Liebend{\"o}rfer}
  et~al.}(2008)\citenamefont{{Liebend{\"o}rfer}, {Fischer}, {Fr{\"o}hlich},
  {Hix}, {Langanke}, {Martinez-Pinedo}, {Mezzacappa}, {Scheidegger},
  {Thielemann}, and {Whitehouse}}}]{liebendorfer-etal:2008}
\bibinfo{author}{\bibfnamefont{M.}~\bibnamefont{{Liebend{\"o}rfer}}},
  \bibinfo{author}{\bibfnamefont{T.}~\bibnamefont{{Fischer}}},
  \bibinfo{author}{\bibfnamefont{C.}~\bibnamefont{{Fr{\"o}hlich}}},
  \bibinfo{author}{\bibfnamefont{W.~R.} \bibnamefont{{Hix}}},
  \bibinfo{author}{\bibfnamefont{K.}~\bibnamefont{{Langanke}}},
  \bibinfo{author}{\bibfnamefont{G.}~\bibnamefont{{Martinez-Pinedo}}},
  \bibinfo{author}{\bibfnamefont{A.}~\bibnamefont{{Mezzacappa}}},
  \bibinfo{author}{\bibfnamefont{S.}~\bibnamefont{{Scheidegger}}},
  \bibinfo{author}{\bibfnamefont{F.-K.} \bibnamefont{{Thielemann}}},
  \bibnamefont{and} \bibinfo{author}{\bibfnamefont{S.~C.}
  \bibnamefont{{Whitehouse}}}, \bibinfo{journal}{New Astronomy Reviews}
  \textbf{\bibinfo{volume}{52}}, \bibinfo{pages}{373} (\bibinfo{year}{2008}).

\bibitem[{\citenamefont{{Liebend{\"o}rfer}
  et~al.}(2009)\citenamefont{{Liebend{\"o}rfer}, {Fischer}, {Hempel},
  {Mezzacappa}, {Pagliara}, {Sagert}, {Schaffner-Bielich}, {Scheidegger},
  {Thielemann}, and {Whitehouse}}}]{liebendorfer-etal:2009}
\bibinfo{author}{\bibfnamefont{M.}~\bibnamefont{{Liebend{\"o}rfer}}},
  \bibinfo{author}{\bibfnamefont{T.}~\bibnamefont{{Fischer}}},
  \bibinfo{author}{\bibfnamefont{M.}~\bibnamefont{{Hempel}}},
  \bibinfo{author}{\bibfnamefont{A.}~\bibnamefont{{Mezzacappa}}},
  \bibinfo{author}{\bibfnamefont{G.}~\bibnamefont{{Pagliara}}},
  \bibinfo{author}{\bibfnamefont{I.}~\bibnamefont{{Sagert}}},
  \bibinfo{author}{\bibfnamefont{J.}~\bibnamefont{{Schaffner-Bielich}}},
  \bibinfo{author}{\bibfnamefont{S.}~\bibnamefont{{Scheidegger}}},
  \bibinfo{author}{\bibfnamefont{F.-K.} \bibnamefont{{Thielemann}}},
  \bibnamefont{and} \bibinfo{author}{\bibfnamefont{S.~C.}
  \bibnamefont{{Whitehouse}}}, \bibinfo{journal}{Nuclear Physics A}
  \textbf{\bibinfo{volume}{827}}, \bibinfo{pages}{573} (\bibinfo{year}{2009}).

\bibitem[{\citenamefont{{Hammer} et~al.}(2010)\citenamefont{{Hammer}, {Janka},
  and {M{\"u}ller}}}]{hjm:2010}
\bibinfo{author}{\bibfnamefont{N.~J.} \bibnamefont{{Hammer}}},
  \bibinfo{author}{\bibfnamefont{H.-T.} \bibnamefont{{Janka}}},
  \bibnamefont{and}
  \bibinfo{author}{\bibfnamefont{E.}~\bibnamefont{{M{\"u}ller}}},
  \bibinfo{journal}{\apj} \textbf{\bibinfo{volume}{714}}, \bibinfo{pages}{1371}
  (\bibinfo{year}{2010}), \eprint{0908.3474}.

\bibitem[{\citenamefont{{Hix} et~al.}(2010)\citenamefont{{Hix}, {Lentz},
  {Baird}, {Messer}, {Mezzacappa}, {Lee}, {Bruenn}, {Blondin}, and
  {Marronetti}}}]{hix-etal:2010}
\bibinfo{author}{\bibfnamefont{W.~R.} \bibnamefont{{Hix}}},
  \bibinfo{author}{\bibfnamefont{E.~J.} \bibnamefont{{Lentz}}},
  \bibinfo{author}{\bibfnamefont{M.}~\bibnamefont{{Baird}}},
  \bibinfo{author}{\bibfnamefont{O.~E.~B.} \bibnamefont{{Messer}}},
  \bibinfo{author}{\bibfnamefont{A.}~\bibnamefont{{Mezzacappa}}},
  \bibinfo{author}{\bibfnamefont{C.-T.} \bibnamefont{{Lee}}},
  \bibinfo{author}{\bibfnamefont{S.~W.} \bibnamefont{{Bruenn}}},
  \bibinfo{author}{\bibfnamefont{J.~M.} \bibnamefont{{Blondin}}},
  \bibnamefont{and}
  \bibinfo{author}{\bibfnamefont{P.}~\bibnamefont{{Marronetti}}},
  \bibinfo{journal}{Nuclear Physics A} \textbf{\bibinfo{volume}{834}},
  \bibinfo{pages}{602} (\bibinfo{year}{2010}).

\bibitem[{\citenamefont{{Burrows} et~al.}(2012)\citenamefont{{Burrows},
  {Dolence}, and {Murphy}}}]{bdm:2012}
\bibinfo{author}{\bibfnamefont{A.}~\bibnamefont{{Burrows}}},
  \bibinfo{author}{\bibfnamefont{J.~C.} \bibnamefont{{Dolence}}},
  \bibnamefont{and} \bibinfo{author}{\bibfnamefont{J.~W.}
  \bibnamefont{{Murphy}}}, \bibinfo{journal}{ArXiv e-prints}
  (\bibinfo{year}{2012}), \eprint{1204.3088}.

\bibitem[{\citenamefont{{Bethe} et~al.}(1979)\citenamefont{{Bethe}, {Brown},
  {Applegate}, and {Lattimer}}}]{bbal:1979}
\bibinfo{author}{\bibfnamefont{H.~A.} \bibnamefont{{Bethe}}},
  \bibinfo{author}{\bibfnamefont{G.~E.} \bibnamefont{{Brown}}},
  \bibinfo{author}{\bibfnamefont{J.}~\bibnamefont{{Applegate}}},
  \bibnamefont{and} \bibinfo{author}{\bibfnamefont{J.~M.}
  \bibnamefont{{Lattimer}}}, \bibinfo{journal}{Nuclear Physics A}
  \textbf{\bibinfo{volume}{324}}, \bibinfo{pages}{487} (\bibinfo{year}{1979}).

\bibitem[{\citenamefont{{Fuller}}(1982)}]{fuller:1982}
\bibinfo{author}{\bibfnamefont{G.~M.} \bibnamefont{{Fuller}}},
  \bibinfo{journal}{\apj} \textbf{\bibinfo{volume}{252}}, \bibinfo{pages}{741}
  (\bibinfo{year}{1982}).

\bibitem[{\citenamefont{{Bahcall} et~al.}(1974)\citenamefont{{Bahcall},
  {Treiman}, and {Zee}}}]{btz:1974}
\bibinfo{author}{\bibfnamefont{J.~N.} \bibnamefont{{Bahcall}}},
  \bibinfo{author}{\bibfnamefont{S.~B.} \bibnamefont{{Treiman}}},
  \bibnamefont{and} \bibinfo{author}{\bibfnamefont{A.}~\bibnamefont{{Zee}}},
  \bibinfo{journal}{Physics Letters B} \textbf{\bibinfo{volume}{52}},
  \bibinfo{pages}{275} (\bibinfo{year}{1974}).

\bibitem[{\citenamefont{{Dicus} et~al.}(1976)\citenamefont{{Dicus}, {Kolb},
  {Schramm}, and {Tubbs}}}]{dkst:1976}
\bibinfo{author}{\bibfnamefont{D.~A.} \bibnamefont{{Dicus}}},
  \bibinfo{author}{\bibfnamefont{E.~W.} \bibnamefont{{Kolb}}},
  \bibinfo{author}{\bibfnamefont{D.~N.} \bibnamefont{{Schramm}}},
  \bibnamefont{and} \bibinfo{author}{\bibfnamefont{D.~L.}
  \bibnamefont{{Tubbs}}}, \bibinfo{journal}{\apj}
  \textbf{\bibinfo{volume}{210}}, \bibinfo{pages}{481} (\bibinfo{year}{1976}).

\bibitem[{\citenamefont{{Fuller} and {Meyer}}(1991)}]{fm:1991}
\bibinfo{author}{\bibfnamefont{G.~M.} \bibnamefont{{Fuller}}} \bibnamefont{and}
  \bibinfo{author}{\bibfnamefont{B.~S.} \bibnamefont{{Meyer}}},
  \bibinfo{journal}{\apj} \textbf{\bibinfo{volume}{376}}, \bibinfo{pages}{701}
  (\bibinfo{year}{1991}).

\bibitem[{\citenamefont{{Misch} et~al.}(2013)\citenamefont{{Misch}, {Brown},
  and {Fuller}}}]{mbf:2013}
\bibinfo{author}{\bibfnamefont{G.~W.} \bibnamefont{{Misch}}},
  \bibinfo{author}{\bibfnamefont{B.~A.} \bibnamefont{{Brown}}},
  \bibnamefont{and} \bibinfo{author}{\bibfnamefont{G.~M.}
  \bibnamefont{{Fuller}}}, \bibinfo{journal}{\prc}
  \textbf{\bibinfo{volume}{88}}, \bibinfo{eid}{015807} (\bibinfo{year}{2013}),
  \eprint{1301.7042}.

\bibitem[{\citenamefont{{Bethe}}(1936)}]{bethe:1936}
\bibinfo{author}{\bibfnamefont{H.~A.} \bibnamefont{{Bethe}}},
  \bibinfo{journal}{Physical Review} \textbf{\bibinfo{volume}{50}},
  \bibinfo{pages}{332} (\bibinfo{year}{1936}).

\bibitem[{\citenamefont{Firestone}(1996)}]{toi}
\bibinfo{author}{\bibfnamefont{R.~B.} \bibnamefont{Firestone}},
  \emph{\bibinfo{title}{Table of Isotopes}}
  (\bibinfo{publisher}{Wiley-Interscience}, \bibinfo{year}{1996}),
  \bibinfo{edition}{1st} ed.

\bibitem[{\citenamefont{{Yako} et~al.}(2005)\citenamefont{{Yako}, {Sakai},
  {Greenfield}, {Hatanaka}, {Hatano}, {Kamiya}, {Kato}, {Kitamura}, {Maeda},
  {Morris} et~al.}}]{yako-etal:2005}
\bibinfo{author}{\bibfnamefont{K.}~\bibnamefont{{Yako}}},
  \bibinfo{author}{\bibfnamefont{H.}~\bibnamefont{{Sakai}}},
  \bibinfo{author}{\bibfnamefont{M.~B.} \bibnamefont{{Greenfield}}},
  \bibinfo{author}{\bibfnamefont{K.}~\bibnamefont{{Hatanaka}}},
  \bibinfo{author}{\bibfnamefont{M.}~\bibnamefont{{Hatano}}},
  \bibinfo{author}{\bibfnamefont{J.}~\bibnamefont{{Kamiya}}},
  \bibinfo{author}{\bibfnamefont{H.}~\bibnamefont{{Kato}}},
  \bibinfo{author}{\bibfnamefont{Y.}~\bibnamefont{{Kitamura}}},
  \bibinfo{author}{\bibfnamefont{Y.}~\bibnamefont{{Maeda}}},
  \bibinfo{author}{\bibfnamefont{C.~L.} \bibnamefont{{Morris}}},
  \bibnamefont{et~al.}, \bibinfo{journal}{Physics Letters B}
  \textbf{\bibinfo{volume}{615}}, \bibinfo{pages}{193} (\bibinfo{year}{2005}),
  \eprint{nucl-ex/0411011}.

\bibitem[{\citenamefont{{Yako} et~al.}(2009)\citenamefont{{Yako}, {Sasano},
  {Miki}, {Sakai}, {Dozono}, {Frekers}, {Greenfield}, {Hatanaka}, {Ihara},
  {Kato} et~al.}}]{yako-etal:2009}
\bibinfo{author}{\bibfnamefont{K.}~\bibnamefont{{Yako}}},
  \bibinfo{author}{\bibfnamefont{M.}~\bibnamefont{{Sasano}}},
  \bibinfo{author}{\bibfnamefont{K.}~\bibnamefont{{Miki}}},
  \bibinfo{author}{\bibfnamefont{H.}~\bibnamefont{{Sakai}}},
  \bibinfo{author}{\bibfnamefont{M.}~\bibnamefont{{Dozono}}},
  \bibinfo{author}{\bibfnamefont{D.}~\bibnamefont{{Frekers}}},
  \bibinfo{author}{\bibfnamefont{M.~B.} \bibnamefont{{Greenfield}}},
  \bibinfo{author}{\bibfnamefont{K.}~\bibnamefont{{Hatanaka}}},
  \bibinfo{author}{\bibfnamefont{E.}~\bibnamefont{{Ihara}}},
  \bibinfo{author}{\bibfnamefont{M.}~\bibnamefont{{Kato}}},
  \bibnamefont{et~al.}, \bibinfo{journal}{Physical Review Letters}
  \textbf{\bibinfo{volume}{103}}, \bibinfo{eid}{012503} (\bibinfo{year}{2009}).

\bibitem[{\citenamefont{{Frekers} et~al.}(2013)\citenamefont{{Frekers},
  {Puppe}, {Thies}, and {Ejiri}}}]{fpte:2013}
\bibinfo{author}{\bibfnamefont{D.}~\bibnamefont{{Frekers}}},
  \bibinfo{author}{\bibfnamefont{P.}~\bibnamefont{{Puppe}}},
  \bibinfo{author}{\bibfnamefont{J.~H.} \bibnamefont{{Thies}}},
  \bibnamefont{and} \bibinfo{author}{\bibfnamefont{H.}~\bibnamefont{{Ejiri}}},
  \bibinfo{journal}{Nuclear Physics A} \textbf{\bibinfo{volume}{916}},
  \bibinfo{pages}{219} (\bibinfo{year}{2013}).

\bibitem[{\citenamefont{{Extreme Light Infrastructure}}()}]{eli}
\bibinfo{author}{\bibnamefont{{Extreme Light Infrastructure}}},
  \urlprefix\url{http://www.eli-laser.eu/}.

\bibitem[{\citenamefont{{Fuller}
  et~al.}(1982{\natexlab{a}})\citenamefont{{Fuller}, {Fowler}, and
  {Newman}}}]{ffn:1982a}
\bibinfo{author}{\bibfnamefont{G.~M.} \bibnamefont{{Fuller}}},
  \bibinfo{author}{\bibfnamefont{W.~A.} \bibnamefont{{Fowler}}},
  \bibnamefont{and} \bibinfo{author}{\bibfnamefont{M.~J.}
  \bibnamefont{{Newman}}}, \bibinfo{journal}{\apj}
  \textbf{\bibinfo{volume}{252}}, \bibinfo{pages}{715}
  (\bibinfo{year}{1982}{\natexlab{a}}).

\bibitem[{\citenamefont{{Fuller}
  et~al.}(1982{\natexlab{b}})\citenamefont{{Fuller}, {Fowler}, and
  {Newman}}}]{ffn:1982b}
\bibinfo{author}{\bibfnamefont{G.~M.} \bibnamefont{{Fuller}}},
  \bibinfo{author}{\bibfnamefont{W.~A.} \bibnamefont{{Fowler}}},
  \bibnamefont{and} \bibinfo{author}{\bibfnamefont{M.~J.}
  \bibnamefont{{Newman}}}, \bibinfo{journal}{Astrophys. J. (Supplement)}
  \textbf{\bibinfo{volume}{48}}, \bibinfo{pages}{279}
  (\bibinfo{year}{1982}{\natexlab{b}}).

\bibitem[{\citenamefont{{Fuller} et~al.}(1985)\citenamefont{{Fuller}, {Fowler},
  and {Newman}}}]{ffn:1985}
\bibinfo{author}{\bibfnamefont{G.~M.} \bibnamefont{{Fuller}}},
  \bibinfo{author}{\bibfnamefont{W.~A.} \bibnamefont{{Fowler}}},
  \bibnamefont{and} \bibinfo{author}{\bibfnamefont{M.~J.}
  \bibnamefont{{Newman}}}, \bibinfo{journal}{\apj}
  \textbf{\bibinfo{volume}{293}}, \bibinfo{pages}{1} (\bibinfo{year}{1985}).

\bibitem[{\citenamefont{{Angell} et~al.}(2012)\citenamefont{{Angell},
  {Hammond}, {Karwowski}, {Kelley}, {Krti{\v c}ka}, {Kwan}, {Makinaga}, and
  {Rusev}}}]{angell-etal:2012}
\bibinfo{author}{\bibfnamefont{C.~T.} \bibnamefont{{Angell}}},
  \bibinfo{author}{\bibfnamefont{S.~L.} \bibnamefont{{Hammond}}},
  \bibinfo{author}{\bibfnamefont{H.~J.} \bibnamefont{{Karwowski}}},
  \bibinfo{author}{\bibfnamefont{J.~H.} \bibnamefont{{Kelley}}},
  \bibinfo{author}{\bibfnamefont{M.}~\bibnamefont{{Krti{\v c}ka}}},
  \bibinfo{author}{\bibfnamefont{E.}~\bibnamefont{{Kwan}}},
  \bibinfo{author}{\bibfnamefont{A.}~\bibnamefont{{Makinaga}}},
  \bibnamefont{and} \bibinfo{author}{\bibfnamefont{G.}~\bibnamefont{{Rusev}}},
  \bibinfo{journal}{\prc} \textbf{\bibinfo{volume}{86}}, \bibinfo{eid}{051302}
  (\bibinfo{year}{2012}).

\bibitem[{\citenamefont{{Nabi} and {Sajjad}}(2007)}]{ns:2007}
\bibinfo{author}{\bibfnamefont{J.-U.} \bibnamefont{{Nabi}}} \bibnamefont{and}
  \bibinfo{author}{\bibfnamefont{M.}~\bibnamefont{{Sajjad}}},
  \bibinfo{journal}{\prc} \textbf{\bibinfo{volume}{76}}, \bibinfo{eid}{055803}
  (\bibinfo{year}{2007}), \eprint{1108.0819}.

\bibitem[{\citenamefont{{Nabi}}(2011)}]{nabi:2011}
\bibinfo{author}{\bibfnamefont{J.-U.} \bibnamefont{{Nabi}}},
  \bibinfo{journal}{Advances in Space Research} \textbf{\bibinfo{volume}{48}},
  \bibinfo{pages}{985} (\bibinfo{year}{2011}), \eprint{1203.4344}.

\bibitem[{\citenamefont{{Nabi}}(2012)}]{nabi:2012}
\bibinfo{author}{\bibfnamefont{J.-U.} \bibnamefont{{Nabi}}},
  \bibinfo{journal}{European Physical Journal A} \textbf{\bibinfo{volume}{48}},
  \bibinfo{pages}{84} (\bibinfo{year}{2012}).

\bibitem[{\citenamefont{{Oda} et~al.}(1994)\citenamefont{{Oda}, {Hino}, {Muto},
  {Takahara}, and {Sato}}}]{oda-etal:1994}
\bibinfo{author}{\bibfnamefont{T.}~\bibnamefont{{Oda}}},
  \bibinfo{author}{\bibfnamefont{M.}~\bibnamefont{{Hino}}},
  \bibinfo{author}{\bibfnamefont{K.}~\bibnamefont{{Muto}}},
  \bibinfo{author}{\bibfnamefont{M.}~\bibnamefont{{Takahara}}},
  \bibnamefont{and} \bibinfo{author}{\bibfnamefont{K.}~\bibnamefont{{Sato}}},
  \bibinfo{journal}{Atomic Data and Nuclear Data Tables}
  \textbf{\bibinfo{volume}{56}}, \bibinfo{pages}{231} (\bibinfo{year}{1994}).

\bibitem[{\citenamefont{{Dzhioev} et~al.}(2010)\citenamefont{{Dzhioev},
  {Vdovin}, {Ponomarev}, {Wambach}, {Langanke}, and
  {Mart{\'{\i}}nez-Pinedo}}}]{dzhioev-etal:2010}
\bibinfo{author}{\bibfnamefont{A.~A.} \bibnamefont{{Dzhioev}}},
  \bibinfo{author}{\bibfnamefont{A.~I.} \bibnamefont{{Vdovin}}},
  \bibinfo{author}{\bibfnamefont{V.~Y.} \bibnamefont{{Ponomarev}}},
  \bibinfo{author}{\bibfnamefont{J.}~\bibnamefont{{Wambach}}},
  \bibinfo{author}{\bibfnamefont{K.}~\bibnamefont{{Langanke}}},
  \bibnamefont{and}
  \bibinfo{author}{\bibfnamefont{G.}~\bibnamefont{{Mart{\'{\i}}nez-Pinedo}}},
  \bibinfo{journal}{\prc} \textbf{\bibinfo{volume}{81}}, \bibinfo{eid}{015804}
  (\bibinfo{year}{2010}), \eprint{0911.0303}.

\bibitem[{\citenamefont{{Sarriguren}}(2013)}]{sarriguren:2013}
\bibinfo{author}{\bibfnamefont{P.}~\bibnamefont{{Sarriguren}}},
  \bibinfo{journal}{\prc} \textbf{\bibinfo{volume}{87}}, \bibinfo{eid}{045801}
  (\bibinfo{year}{2013}), \eprint{1304.2155}.

\bibitem[{\citenamefont{{Brown} et~al.}(2004)\citenamefont{{Brown},
  {Etchegoyen}, {Godwin}, {Rae}, {Richter}, {Ormand}, {Warburton}, {Winfield},
  {Zhao}, and {Zimmerman}}}]{oxbash}
\bibinfo{author}{\bibfnamefont{B.~A.} \bibnamefont{{Brown}}},
  \bibinfo{author}{\bibfnamefont{A.}~\bibnamefont{{Etchegoyen}}},
  \bibinfo{author}{\bibfnamefont{N.~S.} \bibnamefont{{Godwin}}},
  \bibinfo{author}{\bibfnamefont{W.~D.~M.} \bibnamefont{{Rae}}},
  \bibinfo{author}{\bibfnamefont{W.}~\bibnamefont{{Richter}}},
  \bibinfo{author}{\bibfnamefont{W.~E.} \bibnamefont{{Ormand}}},
  \bibinfo{author}{\bibfnamefont{E.~K.} \bibnamefont{{Warburton}}},
  \bibinfo{author}{\bibfnamefont{J.~S.} \bibnamefont{{Winfield}}},
  \bibinfo{author}{\bibfnamefont{L.}~\bibnamefont{{Zhao}}}, \bibnamefont{and}
  \bibinfo{author}{\bibfnamefont{C.~H.} \bibnamefont{{Zimmerman}}},
  \bibinfo{journal}{MSU-NSCL Report No. 1289}  (\bibinfo{year}{2004}).

\bibitem[{\citenamefont{{Brown} and {Richter}}(2006)}]{br:2006}
\bibinfo{author}{\bibfnamefont{B.~A.} \bibnamefont{{Brown}}} \bibnamefont{and}
  \bibinfo{author}{\bibfnamefont{W.~A.} \bibnamefont{{Richter}}},
  \bibinfo{journal}{\prc} \textbf{\bibinfo{volume}{74}}, \bibinfo{eid}{034315}
  (\bibinfo{year}{2006}).

\bibitem[{\citenamefont{{Brown} and {Wildenthal}}(1985)}]{brownwildenthal:1985}
\bibinfo{author}{\bibfnamefont{B.~A.} \bibnamefont{{Brown}}} \bibnamefont{and}
  \bibinfo{author}{\bibfnamefont{B.~H.} \bibnamefont{{Wildenthal}}},
  \bibinfo{journal}{Atomic Data and Nuclear Data Tables}
  \textbf{\bibinfo{volume}{33}}, \bibinfo{pages}{347} (\bibinfo{year}{1985}).

\bibitem[{\citenamefont{{Frazier} et~al.}(1997)\citenamefont{{Frazier},
  {Brown}, {Millener}, and {Zelevinsky}}}]{fbmz:1997}
\bibinfo{author}{\bibfnamefont{N.}~\bibnamefont{{Frazier}}},
  \bibinfo{author}{\bibfnamefont{B.~A.} \bibnamefont{{Brown}}},
  \bibinfo{author}{\bibfnamefont{D.~J.} \bibnamefont{{Millener}}},
  \bibnamefont{and}
  \bibinfo{author}{\bibfnamefont{V.}~\bibnamefont{{Zelevinsky}}},
  \bibinfo{journal}{Physics Letters B} \textbf{\bibinfo{volume}{414}},
  \bibinfo{pages}{7} (\bibinfo{year}{1997}).

\end{thebibliography}

\end{document}